\documentclass{emulateapj}
\usepackage{graphics}
\usepackage{lscape, graphicx}
\usepackage{natbib}

\newcommand{\etal}{et al.}  
\newcommand{\per}{\ensuremath{^{-1}}}
\newcommand{\persq}{\ensuremath{^{-2}}}

\newcommand{\hal}{H\ensuremath{\alpha}}
\newcommand{\hbeta}{H\ensuremath{\beta}} 
\newcommand{\hst}{\emph{HST}}
\newcommand{\msun}{\ensuremath{M_{\odot}}}
\newcommand{\kms}{km~s\ensuremath{^{-1}}} 
\newcommand{\ergs}{ergs~s\ensuremath{^{-1}}}

\newcommand{\ergcmsH}{ergs~cm\ensuremath{^{-2}}~s\ensuremath{^{-1}}~Hz\ensuremath{^{-1}}}
\newcommand{\ergsH}{ergs~s\ensuremath{^{-1}}~Hz\ensuremath{^{-1}}}

\newcommand{\mbh}{\ensuremath{M_\mathrm{BH}}}

\newcommand{\sigmastar}{\ensuremath{\sigma_{*}}}
\newcommand{\mbul}{\ensuremath{M_\mathrm{bul}}}
\newcommand{\msigma}{\ensuremath{\mbh-\sigma_{*}}}
\newcommand{\mlbul}{\ensuremath{\mbh-\lbul}}
\newcommand{\rblr}{\ensuremath{r_{\mathrm{BLR}}}}
\newcommand{\vfwhm}{\ensuremath{v_{\mathrm{FWHM}}}}

\newcommand{\xmmn}{\emph{XMM-Newton}}
\newcommand{\chandra}{\emph{Chandra}}

\newcommand{\lbol}{\ensuremath{L_{\mathrm{bol}}}}
\newcommand{\lbul}{\ensuremath{L_{\mathrm{bul}}}}
\newcommand{\ledd}{\ensuremath{L_{\mathrm{Edd}}}}
\newcommand{\alOX}{\ensuremath{\alpha_{\mathrm{OX}}}}
\newcommand{\sersic}{S\'{e}rsic}

\newcommand{\um}{\ensuremath{\mu}m}

\shorttitle{POX 52} 
\shortauthors{THORNTON ET AL.}

\begin{document}
\title{The Host Galaxy and Central Engine of the Dwarf AGN POX 52\altaffilmark{1,2}} 
\author {Carol E. Thornton\altaffilmark{3}, Aaron J. Barth\altaffilmark{3}, Luis C. 
Ho\altaffilmark{4}, Robert E. Rutledge\altaffilmark{5}, and Jenny E. Greene\altaffilmark{6,7}.}

\altaffiltext{1}{Based on observations made with the NASA/ESA Hubble Space Telescope, obtained 
at the Space Telescope Science Institute, which is operated by the Association of Universities 
for Research in Astronomy, Inc., under NASA contract NAS 5-26555. These observations are 
associated with program \# 10239.}

\altaffiltext{2}{Based on observations obtained with XMM-Newton, an ESA science mission with 
instruments and contributions directly funded by ESA Member States and NASA.}

\altaffiltext{3}{Department of Physics \& Astronomy, 4129 Fredrick Reines Hall, University of 
California, Irvine, CA 92619-4575; thorntoc@uci.edu, barth@uci.edu}
\altaffiltext{4}{The Observatories of the Carnegie Institution of Washington, 813 Santa Barbara 
Street, Pasadena, CA 91101}
\altaffiltext{5}{Department of Physics, McGill University, 3600 rue University, Montreal, QC, 
H3A 2T8, Canada} 
\altaffiltext{6}{Department of Astrophysical Sciences, Princeton University, Princeton, NJ 
08544}
\altaffiltext{7}{Hubble Fellow and Princeton-Carnegie Fellow}

\begin{abstract}
We present new multi-wavelength observations of the dwarf Seyfert 1 galaxy POX~52 in order to 
investigate the properties of the host galaxy and the active nucleus, and to examine the mass of
its black hole, previously estimated to be $\sim 10^{5}$ \msun. {\it Hubble Space Telescope} 
ACS/HRC images show that the host galaxy has a dwarf elliptical morphology 
($M_I= -18.4$ mag, \sersic\ index $n=4.3$) with no detected disk component or spiral structure, 
confirming previous results from ground-based imaging. X-ray observations from both \chandra\ 
and \xmmn\ show strong (factor of 2) variability over timescales as short as 500 s, as 
well as a dramatic decrease in the absorbing column density over a 9 month period. We attribute
this change to a partial covering absorber, with a 94\% covering fraction and 
$N_{H} = 58^{+8.4}_{-9.2} \times 10^{21}$ cm$^{-2}$, that moved out of the line of sight in 
between the \xmmn\ and \chandra\ observations. Combining these data with observations from the 
VLA, {\it Spitzer}, and archival data from 2MASS and {\it GALEX}, we examine the spectral energy
distribution (SED) of the active nucleus.  Its shape is broadly similar to typical radio-quiet 
quasar SEDs, despite the very low bolometric luminosity of $\lbol\ = 1.3 \times 10^{43}$ \ergs.
Finally, we compare black hole mass estimators including methods based on X-ray variability, and
optical scaling relations using the broad \hbeta\ line width and AGN continuum luminosity, 
finding a range of black hole mass from all methods to be 
$\mbh\ = (2.2-4.2) \times\ 10^{5}$ \msun, with an Eddington ratio of 
$\lbol/\ledd \approx 0.2-0.5$.
\end{abstract}


\section{Introduction}
\begin{figure*}
\begin{center}
\scalebox{0.25}{\includegraphics{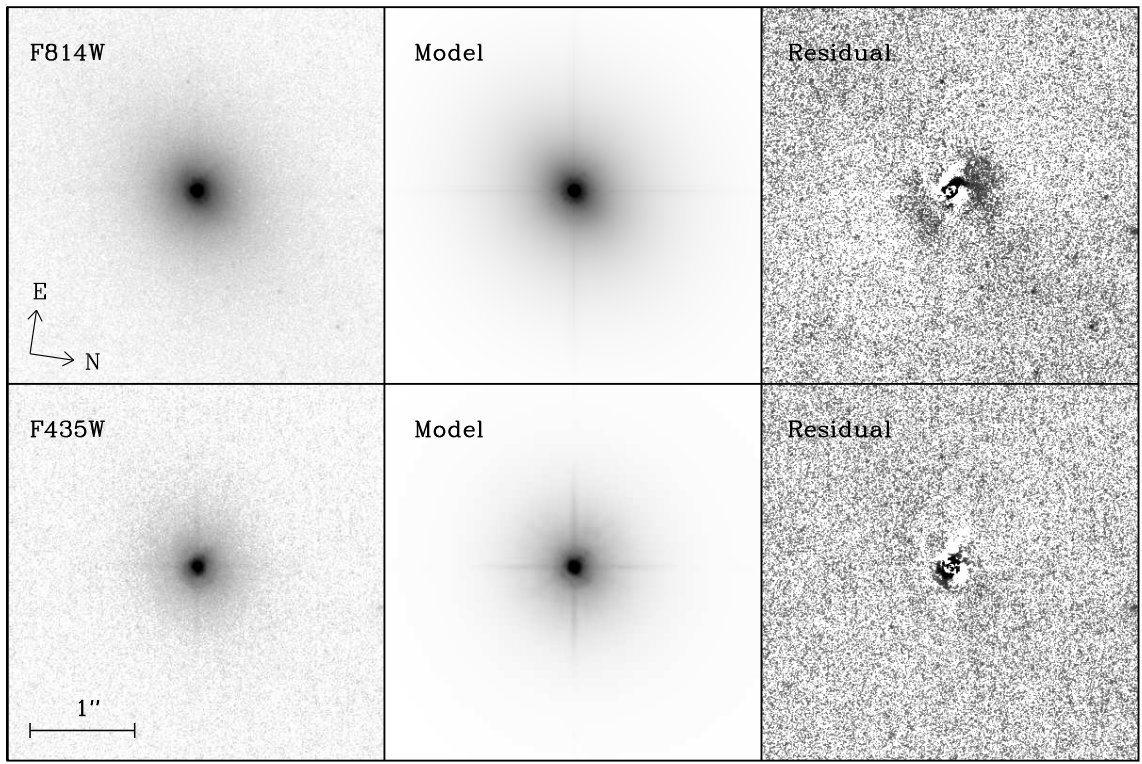}}
\end{center}
\caption{\hst\ ACS/HRC images of POX 52 and GALFIT models. The top and bottom rows show the 
F814W and F435W images, respectively. In each row, the left panel is the ACS image, the middle 
panel is the full GALFIT model, and the right panel shows the fit residuals. For the F435W 
filter, the double-\sersic\ host galaxy model is shown.}
\label{image-panel}
\end{figure*}

Stellar and gas dynamical studies have proven to be the most secure method of measuring the 
masses of supermassive black holes in nearby galaxies, and black holes with masses of 
$\sim10^6~$--$~10^9$ \msun\ have now been detected in a few dozen galaxies (see Ferrarese \&
Ford 2005 for a recent review). With just a few exceptions, however, stellar-dynamical and 
gas-dynamical techniques for black hole mass measurement cannot be applied to Seyfert galaxies 
and quasars, because most luminous active galaxies are too distant for their black hole's 
gravitational sphere of influence to be resolved. Most black hole mass estimates in broad-lined 
active galactic nuclei (AGNs) are based on indirect methods that rely on scaling relations 
resulting from reverberation-mapping data \citep{Kaspi, bentz06}. Using these methods, it has 
recently become possible to search for active galaxies having extremely low-mass black holes, in
order to extend our understanding of black hole demographics to objects having black hole masses
of $\mbh \lesssim 10^6$ \msun.  

The best nearby example of a black hole in the $<10^6$ \msun\ mass range is that in the dwarf 
Seyfert 1 galaxy NGC~4395 \citep{FilS89,fh03}. NGC~4395 is an essentially bulgeless, late-type
dwarf galaxy with type 1 Seyfert characteristics such as broad emission lines 
\citep{FilS89, FHS93} and rapid X-ray variability \citep{Iwa00, Shih03, Moran05}. Located at a 
distance of only $\sim4.3$~Mpc \citep{thim04}, it is an excellent target for multiwavelength 
observations despite the tiny luminosity of its central engine 
($\lbol \approx 6\times10^{40}$ \ergs; Moran \etal\ 1999). A recent ultraviolet (UV) 
reverberation-mapping measurement found \mbh~$=~(3.6~\pm~1.1)~\times~10^5$~\msun\ \citep{Pet05}.
\citet{GH04, GH07c} searched the Sloan Digital Sky Survey (SDSS) archives to find AGNs 
with low-mass black holes and have found numerous examples of Seyfert 1 galaxies with 
$\mbh < 10^6$ \msun, which we will refer to as the GH sample. However, most of the SDSS objects 
found by Greene \& Ho are much more distant than NGC 4395, making detailed studies of their 
nuclei and host galaxies more difficult. Although the GH sample does contain examples of AGNs in
low-mass disk galaxies \citep[see also][]{Dong}, NGC 4395 remains an exceptional case due to its
extreme late-type morphology as well as its proximity.

\object{POX 52}, also known as G1200-2038 or PGC 038055 
($D~=~93$ Mpc for $H_0= 70$ \kms\ Mpc\per), is an interesting counterpart to NGC~4395. It was 
originally discovered in an objective prism study by \citet{KSB87}, who classified it as a dwarf
disk galaxy with a Seyfert 2 nucleus, but noted the presence of a weak broad component of the 
\hbeta\ line. \citet{bar04} re-observed POX 52, classifying it as a type 1 AGN based on clear 
detections of broad \hal\ and \hbeta\ and noting that its optical emission-line spectrum is 
nearly identical to that of NGC 4395. Both NGC 4395 and POX 52 can be classified as narrow-line 
Seyfert 1 (NLS1) galaxies, since they satisfy the defining criterion of having 
FWHM(\hbeta)$<2000$ \kms. NLS1 galaxies are often, but not always, characterized by strong 
\ion{Fe}{2} emission and relatively weak [\ion{O}{3}] emission when compared to \hbeta\, along 
with steep soft ($0.5 - 2.0$ keV) spectral slopes and rapid variability in the X-ray 
\citep{OP85, Boller96}. NGC~4395 and POX~52 are therefore somewhat atypical NLS1 galaxies in 
that they have very high-excitation spectra with large [\ion{O}{3}]/\hbeta\ ratios and very weak
\ion{Fe}{2} emission. 

Using the virial scaling relation calibrated by \citet{Kaspi}, Barth \etal\ estimated a black 
hole mass of $\mbh\approx1.6\times10^{5}$ \msun, making this one of the least massive black 
holes identified in any AGN. This mass is consistent with the mass expected based on the 
\msigma\ relation (extrapolated to small \sigmastar) for the measured stellar velocity 
dispersion $\sigmastar = 36 \pm 5$ \kms. They also estimated the bolometric luminosity 
(\lbol\ $\approx 2 \times 10^{43}$ \ergs) and the Eddington ratio 
(\lbol/\ledd\ $\approx 0.5 - 1$) for POX~52. Thus, although POX~52 and NGC~4395 have very 
similar black hole masses and emission-line spectra that are nearly identical in appearance, the
AGN in POX~52 is $\sim 300$ times more luminous than that in NGC~4395, which has an estimated 
\lbol/\ledd\ $\approx 10^{-3}$ \citep{Pet05}. \citet{bar04} used ground-based images to study 
the morphology of POX~52 and found that the host galaxy was best fit with a \sersic\ model with 
an index of $n = 3.6 \pm 0.2$, effective radius of $r_{e} = 1\farcs{2}$, and absolute magnitude 
$M_{V} = -17.6$. Combined with \sigmastar\ measured from the spectrum, these results show POX~52
to fall near the dwarf elliptical sequence \citep{Geha03} in the fundamental plane projections 
of \citet{Burstein97}, making it the first example of a dwarf elliptical to host a Seyfert 
nucleus. Although the past few years have seen a large increase in the number of known AGNs with
low-mass black holes, POX~52 remains a valuable target for further study, because of its 
relatively small distance and the fact that it is a less luminous and presumably less massive 
galaxy than the majority of the host galaxies in the GH SDSS sample.

Among nearby dwarf ellipticals, there is little evidence for the presence of central black 
holes.  The closest well-studied example is NGC 205 in the Local Group, and stellar-dynamical 
observations with the \emph{Hubble Space Telescope (HST)} have set an upper limit to the black 
hole mass of $\mbh < 3.8\times10^4$ \msun\ \citep{Val05}, which is below the mass expected from
a simple extrapolation of the \msigma\ relation. For dwarf elliptical galaxies outside the Local
Group, stellar-dynamical observations lack the spatial resolution needed to detect black holes, 
and AGN surveys are the best available alternative for black hole searches. A double nucleus in 
one Virgo dwarf elliptical galaxy has been interpreted as morphological evidence for a stellar 
disk in orbit around a central black hole \citep{deb06}, but aside from NGC~205 there are 
essentially no direct observational constraints on the black hole occupation fraction in dwarf
elliptical galaxies. POX~52 represents a unique opportunity to study, in depth, the unambiguous 
presence of a low-mass central black hole in a nearby dwarf elliptical galaxy.

Here, we present new multiwavelength observations of POX 52, including the first targeted \hst\
and X-ray observations of this unusual object. \hst\ images 
are used to examine the host galaxy morphology; our results confirm the previous conclusions 
that the galaxy is consistent with a dwarf elliptical morphology with a \sersic\ index of 
$\sim 4$. We describe X-ray observations obtained with both the \chandra\ and \xmmn\ X-ray 
Observatories in order to study the temporal and spectral properties of the active nucleus, and 
to constrain its bolometric luminosity and black hole mass. We also include observations from 
the Very Large Array (VLA) and {\it Spitzer} Infrared Spectrometer (IRS), as well as archival 
data from 2MASS and \emph{GALEX}, in order to investigate the spectral energy distribution (SED)
of POX 52 over a wide range in frequency. We use recently updated versions of the broad-line 
virial scaling relations to obtain new estimates of the mass of the black hole, and in 
combination with the multiwavelength SED we determine the bolometric luminosity and Eddington 
ratio for the AGN. In addition to the \hst\ data presented here, we were also awarded time as 
part of the same project to obtain ultraviolet and optical spectra of POX 52 with the Space 
Telescope Imaging Spectrograph (STIS), but STIS failed before the observations were scheduled to
be taken.

\section{Optical Observations}
Observations of POX 52 were made on 2004 November 18 UT with \hst\ using the High Resolution 
Camera (HRC) on ACS. Four images of 590 s each were taken through the F814W ({\it I}-band)
\dataset[ADS/Sa.HST#J8YB10PQQ]\dataset[ADS/Sa.HST#J8YB10010] filter and four images of 622 s 
each were taken through the F435W ({\it B}-band) \dataset[ADS/Sa.HST#J8YB10PVQ] 
\dataset[ADS/Sa.HST#J8YB10020] filter. These images were dithered using a 4-point box pattern 
with 0\farcs15 spacing to enable the removal of cosmic ray hits and bad pixels. An additional 
100 s image was taken in each filter in the event that the nucleus saturated in the longer 
exposures, though this saturation did not occur in any image. The images for each filter were
combined using the Multidrizzle task in PyRAF (version 1.3) to remove cosmic ray hits and to 
correct for the ACS geometric distortion, and the output images were not sky subtracted or 
rotated during this processing. These \hst\ images (Figure \ref{image-panel}) show POX 52 as a 
bright point source centered in a smooth host galaxy with no obvious signs of spiral structure 
in either filter.

The two-dimensional fitting program GALFIT (version 2.1d, Peng et al.\ 2002) was used to model 
the host galaxy morphology. The central point source in the center of the host galaxy was 
modeled by a point spread function (PSF) image, and the host galaxy itself was modeled by a 
\sersic\ component, or combination of \sersic\ components. Since the sky background was not 
subtracted in the image processing, a sky component was also used within each galaxy model. All 
magnitudes are in the Vegamag system calculated using the zeropoints calibrated for 
HRC\footnotemark, which are 25.188 for F435W and 24.861 for F814W, and corrected for Galactic
extinction using the IRAF package {\it synphot} for $E(B-V) = 0.052$ mag \citep{Schlegel}.
\footnotetext{http://www.stsci.edu/hst/acs/analysis/zeropoints}

\begin{figure}
\begin{center}
\rotatebox{-90}{\scalebox{0.35}{\includegraphics{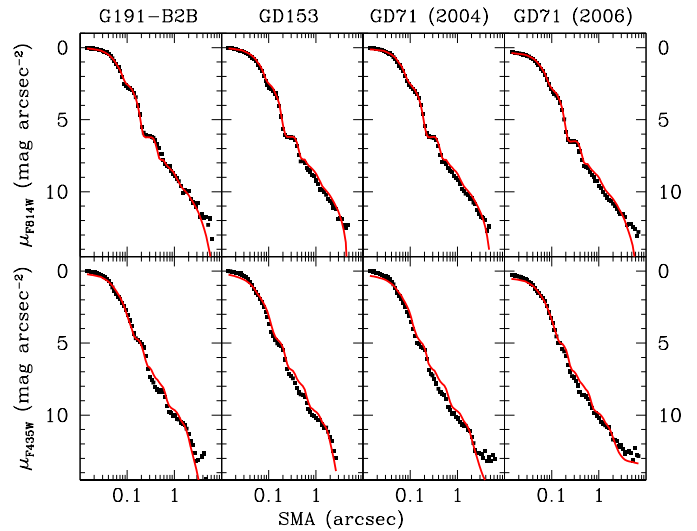}}}
\end{center}
\caption{Radial profile plots of the ACS/HRC standard star images along the semi-major axis 
(SMA), with the Tiny Tim synthetic PSFs over-plotted in red. The magnitude scale is normalized 
to zero mag arcsec$^{-2}$ in the PSF core (except for the 2006 observation of GD71 which uses 
the same offset as the 2004 observation in order to demonstrate time variability of the PSF 
structure). The bump at $\sim~0\farcs4$ in the F814W radial profiles is the the long-wavelength 
scattering halo in ACS/HRC.}
\label{TTmodel}
\end{figure}

\subsection{PSF Modeling}
The PSF image used in the GALFIT modeling can either be created synthetically from a program 
such as Tiny Tim \citep{tinytim} or cropped from a data image using the same detector and filter
set as the object being modeled. Both methods were tested in order to determine the most 
appropriate PSF for our analysis. We began by creating a synthetic PSF for each filter using 
Tiny Tim. We centered the PSF in the same position as the point source in our galaxy images in 
order to properly reproduce the geometric changes involved in the Multidrizzle process. These 
PSF images were then Multidrizzled in the same manner as the object data, since the Tiny Tim 
PSFs created for HRC include the geometric distortions in raw HRC images, and therefore must be 
subject to the same corrections to undistort the images. In order to evaluate how well the Tiny 
Tim PSF compares to the PSF of a star observed with HRC, we searched the archive for unsaturated
images of \hst\ standard stars observed in the F814W and F435W filters, finding three that 
matched our criteria: \object{G191-B2B}, \object{GD153}, and \object{GD71}. Table \ref{PSFstar} 
lists the observation dates and exposure times of these three white dwarf stars, in which they
were located near the center of the CCD. The specific observations were chosen to bracket 
the POX 52 observation in order to minimize any changes to the PSF over time, with the average 
observation occurring 6 months from that of POX 52. All of the PSF star images were processed 
through Multidrizzle in the same manner as our POX 52 data with the exception that the standard 
star images were sky-subtracted during the Multidrizzle process in order to be used as input PSF
images with GALFIT.

\begin{deluxetable}{lllcc}
\tablewidth{0in}
\tablecaption{Stellar PSF Observations}
\tablehead{
\colhead{Star} & \colhead{$V$ (mag)} & \colhead{Obs. Date (UT)} &
\multicolumn{2}{c}{Exposure Time (s)} \\
\colhead{} & \colhead{} & \colhead{} & \colhead{F435W} & \colhead{F814W} }
\startdata
G191-B2B & 11.78 & 2003 Aug 30 & \phn3.2  & \phn4.0 \\
         &       & 2003 Nov 8  & \phn4.0  &    14.0 \\
         &       & 2005 Feb 28 & \phn4.0  &    10.0 \\
GD153    & 13.40 & 2004 Feb 15 &    14.4  &    40.0 \\
         &       & 2005 May 19 &    14.4  &    30.0 \\
GD71     & 13.06 & 2004 Feb 3  &    12.0  &    35.0 \\
         &       & 2006 Nov 20 &    11.0  &    26.0 
\enddata
\label{PSFstar}
\end{deluxetable}

In order to determine how well Tiny Tim models the stellar PSF, we used the Tiny Tim model as 
the input PSF in GALFIT and fit each of the standard stars in both of the observed filters. Not
only can we examine how the shape of the PSF changes for different stars (due to color and/or 
CCD location), but the existence of multiple observations of each of these stars in the \hst\ 
archive allows us to determine the extent of changes to the PSF over time. Figure \ref{TTmodel} 
shows radial profile plots made using the IRAF task {\it ellipse} for the star images compared 
to the GALFIT model using the Tiny Tim PSFs. The three left panels are profiles from the three 
different stars found in the \hst\ archive. One can see how the shape of the PSF's central peak 
can vary depending on the star used. The two observations of GD71 were taken about a year and a 
half apart. There is not a large change in the PSF shape during this time period, but small 
differences are apparent. The differences in profile shapes are more apparent in the F435W 
filter, showing Tiny Tim has more difficulty modeling a PSF in this filter. The Tiny Tim F435W 
PSF tends to have a flatter inner core, leaving a $\sim 0.3$~mag~arcsec$^{-2}$ residual out 
to $\sim 0.05\arcsec$. In the ACS/HRC camera about $10 - 20\%$ of the PSF flux at long 
wavelengths ($\lambda > 7500$~\AA) scatters into a halo surrounding the PSF core making PSF 
modeling more difficult in red passbands \citep{Sirianni}, such as the F814W filter. The halo 
radius increases with wavelength, and Tiny Tim attempts to model its wavelength-dependent 
structure. Despite the red halo (seen as the bump at $\sim 0.3\arcsec$ in Figure \ref{TTmodel}) 
and the variations seen between the stellar PSFs, Tiny Tim did a reasonable job modeling each of
the standard stars in the F814W filter, with few significant deviations between the shape of the
Tiny Tim PSF and that of the observed star. Since the Tiny Tim PSFs reasonably fit the central 
regions of the observed stars in both filters and have the advantage of being noise-free, we 
selected the corresponding Tiny Tim PSF to use in the GALFIT modeling of POX 52.

\subsection{AGN and Host Galaxy Modeling}
\begin{figure}
\epsscale{1}
\plotone{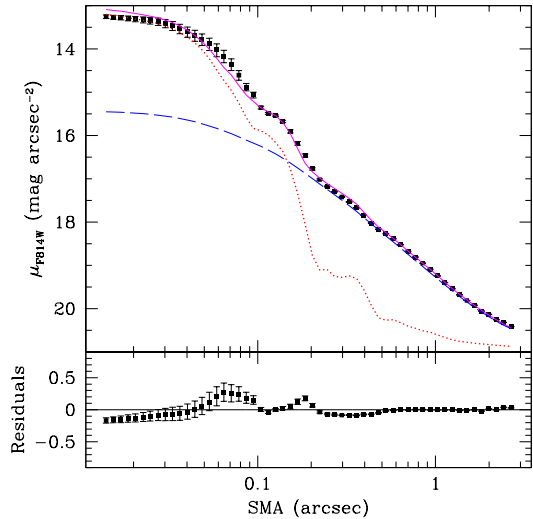}
\caption{Radial profile plot of the surface brightness ({\it top panel}) of POX 52 in the F814W 
filter with residuals ({\it bottom panel}) between the data and the total fit. The black points 
are the data, the solid magenta line is the total fit from GALFIT, the dotted red line is the 
PSF component, and the dashed blue line is the \sersic\ component.}
\label{814_sersic}
\end{figure}

We fit the F814W HRC image of POX 52 over a $\sim 169$ arcsec$^2$ ($491 \times 491$ pixel) 
region with a simple model consisting of an AGN point source and a single \sersic\ component, 
where all parameters (including position and magnitude for both the point source and \sersic\ 
components and effective radius, \sersic\ index, axis ratio, and position angle for the \sersic\
component) were allowed to vary. Figure \ref{814_sersic} shows a radial profile plot of the best
fit from GALFIT with a \sersic\ index of 4.3 and a reduced $\chi^{2}$\ of 15.0 for 241,068 
degrees of freedom (dof). The top panels of Figure \ref{image-panel} show the F814W \hst\ image 
(left panel) and the GALFIT model (middle panel) with 
$m_{\rm AGN} = 17.5$, $m_{\rm host} = 16.2$ and an effective radius of 1\farcs08, or 490 pc. The
residuals after subtracting the model from the galaxy image are shown in the far right panel. 
The strong residuals in the center of the image are mostly confined to the PSF-dominated region 
at $r < 0\farcs1$ and are likely to be the result of PSF mismatch, but the weaker residuals seen
further out from the PSF-dominated region of the galaxy could indicate small-scale structure in 
the host galaxy. We added a second \sersic\ component to the model in an attempt to minimize the
residuals, but it made little difference in the overall fit. The additional \sersic\ component 
closely followed that of the PSF component, leaving the original \sersic\ component to trace the
host galaxy as before. The relative gain in the overall fit did not justify the need for a two 
\sersic\ component model, and, therefore, we select the PSF+\sersic\ model as the best fit the 
F814W images. Parameters for this fit and the F435W fits can be seen in Table \ref{galfitT}. 

\begin{figure}
\epsscale{1}
\plotone{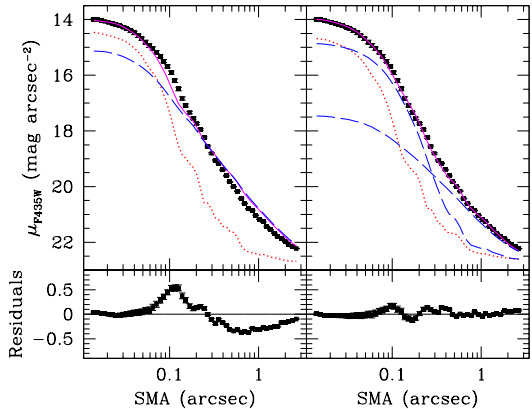}
\caption{Radial profile plots of the surface brightness ({\it top panels}) of POX 52 in the 
F435W filter with residuals ({\it bottom panels}) between the data and the total fit. The black 
points are the data, the solid magenta lines are the total fits from GALFIT, the dotted red 
lines are the PSF components, and the dashed blue lines are \sersic\ components. 
{\it Left Panel:} Best-fit GALFIT model using only one \sersic\ component. 
{\it Right Panel:} Best-fit GALFIT model using two \sersic\ components.}
\label{435_sersic}
\end{figure}

The F435W data proved to be more difficult to model. The galaxy image was initially fit using 
the simple PSF+\sersic\ model as above (Figure \ref{435_sersic}, {\it left panel}), but this 
resulted in an unexpectedly high \sersic\ index of $n = 20$ (the default maximum value for the 
\sersic\ index used by GALFIT) and a poor fit compared with the F814W model results. In order to
get a more reasonable result, we added a second \sersic\ component to the model, allowing the 
parameters for both components to vary freely. In the resulting model fit, the first \sersic\ 
component has a high central surface brightness that traces the central region of the galaxy out
to $\sim 0\farcs25$, resembling the PSF component but with a slightly broader profile, while 
the second \sersic\ component has a much lower surface brightness that models the majority of 
the host galaxy from $\sim 0\farcs5$ outward. This outermost \sersic\ component has an index 
of $n = 4.4$, similar to the result of the F814W fit. 

The effective radius of the host galaxy, as described by the outer \sersic\ component in the 
best-fit (PSF+\sersic+\sersic) model, is slightly larger than in the F814W filter with 
$r_e = 1.22\arcsec = 550$ pc. The reduced $\chi^{2}$ is also larger than in the F814W filter
(26.2 compared to 15.0, even including the second \sersic\ component in the F435W model), and 
the PSF component has an unexpectedly low central surface brightness, as seen in the radial 
profile plot (Figure \ref{435_sersic}, {\it right panel}). 

The inner \sersic\ component consists of some combination of AGN and host galaxy light, and it 
is difficult to disentangle the relative contributions. The magnitude of the PSF component
describes a lower limit to the magnitude of the AGN of $m_{\rm AGN} = 19.2$. We estimate an 
upper limit by repeatedly fitting the galaxy with a PSF+\sersic\ model with a fixed PSF 
magnitude, increasing the magnitude each time until it was bright enough to account for the 
majority of the light in the very center of the galaxy, while allowing the single \sersic\ 
component to vary freely as before. This results in an AGN magnitude of 
$m_{\mathrm{AGN}} = 18.6$, which puts the AGN magnitude in the F435W filter in the range of 
$m_{\rm AGN} = 18.6 - 19.2$ mag. We adopt the magnitude of the outermost \sersic\ component as a
lower limit to the host galaxy brightness, $m_{\rm host} = 18.2$ in the F435W filter and combine
the magnitudes of both \sersic\ components to determine the upper limit, $m_{\rm host} = 17.5$. 
This results in a host galaxy magnitude range of $m_{\rm host} = 17.5 - 18.2$ mag.

\begin{deluxetable}{llllll}
\tablewidth{0in}
\tablecaption{GALFIT Results}
\tablehead{
\colhead{Band} & \colhead{$\chi^{2}/\nu$} & \colhead{Component} & \colhead{Mag} & \colhead{$n$} 
& \colhead{$r_e$ (arcsec/pc)}}
\startdata
F814W                   & 15.0 & PSF        & 17.5 &          &       \\
                        &      & \sersic\   & 16.2 & \phn4.3  & 1.08/490  \\
F435W                   & 31.6 & PSF        & 19.0 &          &       \\
- Single \sersic\       &      & \sersic\   & 16.8 &    20.0  & 5.87/2700  \\
F435W                   & 26.2 & PSF        & 19.2 &          &       \\
- Double \sersic\       &      & \sersic\ 1 & 18.4 & \phn1.4 & 0.060/27 \\
                        &      & \sersic\ 2 & 18.2 & \phn4.4 & 1.22/550  
\enddata
\label{galfitT}
\end{deluxetable}

In order to further confirm our choice of PSF, we refit the galaxy images with both the 
PSF+\sersic\ model and the PSF+\sersic+\sersic\ model using each of the stellar PSFs. In each of
the fits, regardless of the star used or the filter the image was taken in, the PSF component 
drastically under-represented the central point source light contribution. This verified that 
the fits in which the noise-free Tiny Tim PSF was used produced the best overall results.

Our GALFIT results show that the host galaxy of POX 52 has a \sersic\ index $n \approx 4$ in 
both the F814W and F435W filters making it more centrally concentrated than a typical dwarf 
elliptical galaxy with $n \approx 0.3 - 3$ \citep{BJ}. In a study of host galaxies containing 
similar low-mass AGN, \citet{GH08} selected 19 galaxies from the GH sample for \hst\ imaging in 
order to better determine their host galaxy morphology. From this \hst\ imaging, they find that 
60\% of the sample are disk-dominated galaxies, while the remainder are compact systems 
with host morphology similar to POX~52. These compact galaxies have steep \sersic\ indices in 
the range of $2-4$ and reside near dwarf ellipticals from the \citet{GGM}, \citet{Gavazzi}, 
and \citet{MG05} samples in the fundamental plane. Although many of the galaxies in the Greene 
\etal\ \hst\ sample are quite similar to POX~52 in terms of \sersic\ index and overall host 
galaxy morphology, POX~52 ($M_{\rm host} = -18.4$ in the F435W filter) remains less luminous 
than the average galaxy in the GH sample by $\sim 2$ mag \citep{GH08}.

\begin{figure}
\epsscale{1.0}
\plotone{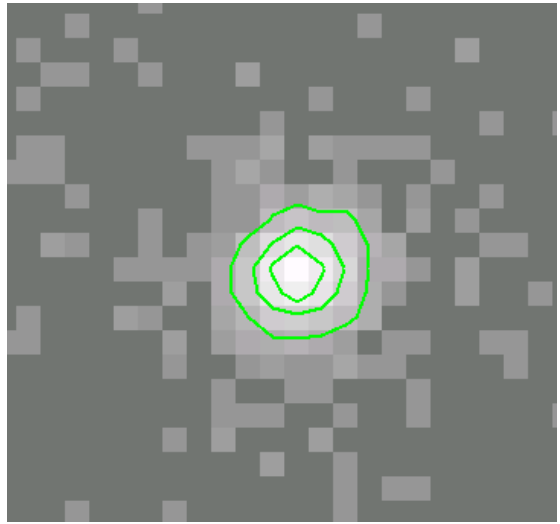}
\caption{The \chandra\ image of POX 52 with intensity (68\%, 90\%, and 99\%) contours overlaid. 
Image is $11\arcsec$ on a side.}
\label{ChandraPS}
\end{figure}

\begin{figure*}
\epsscale{1.0}
\plottwo{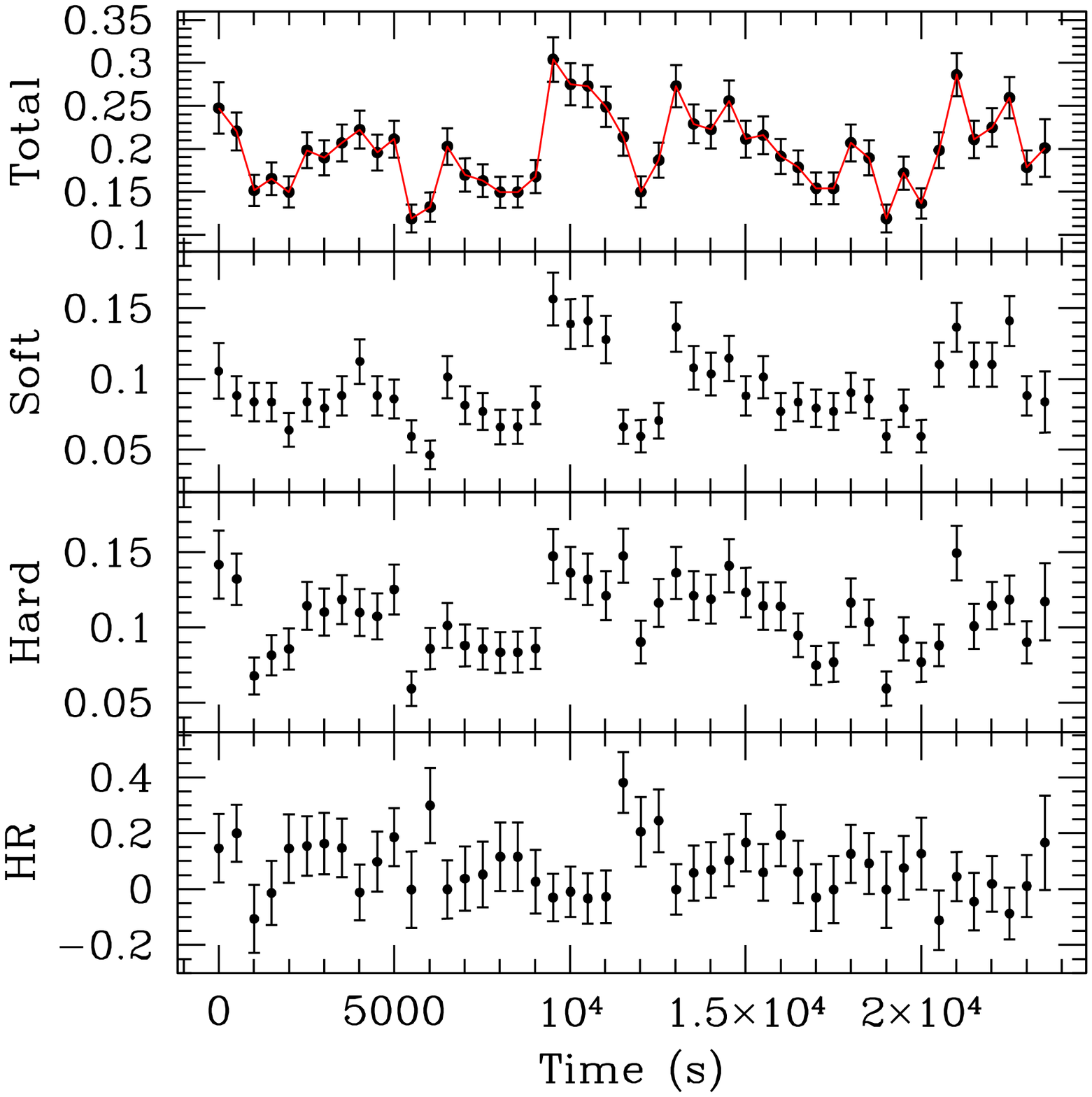}{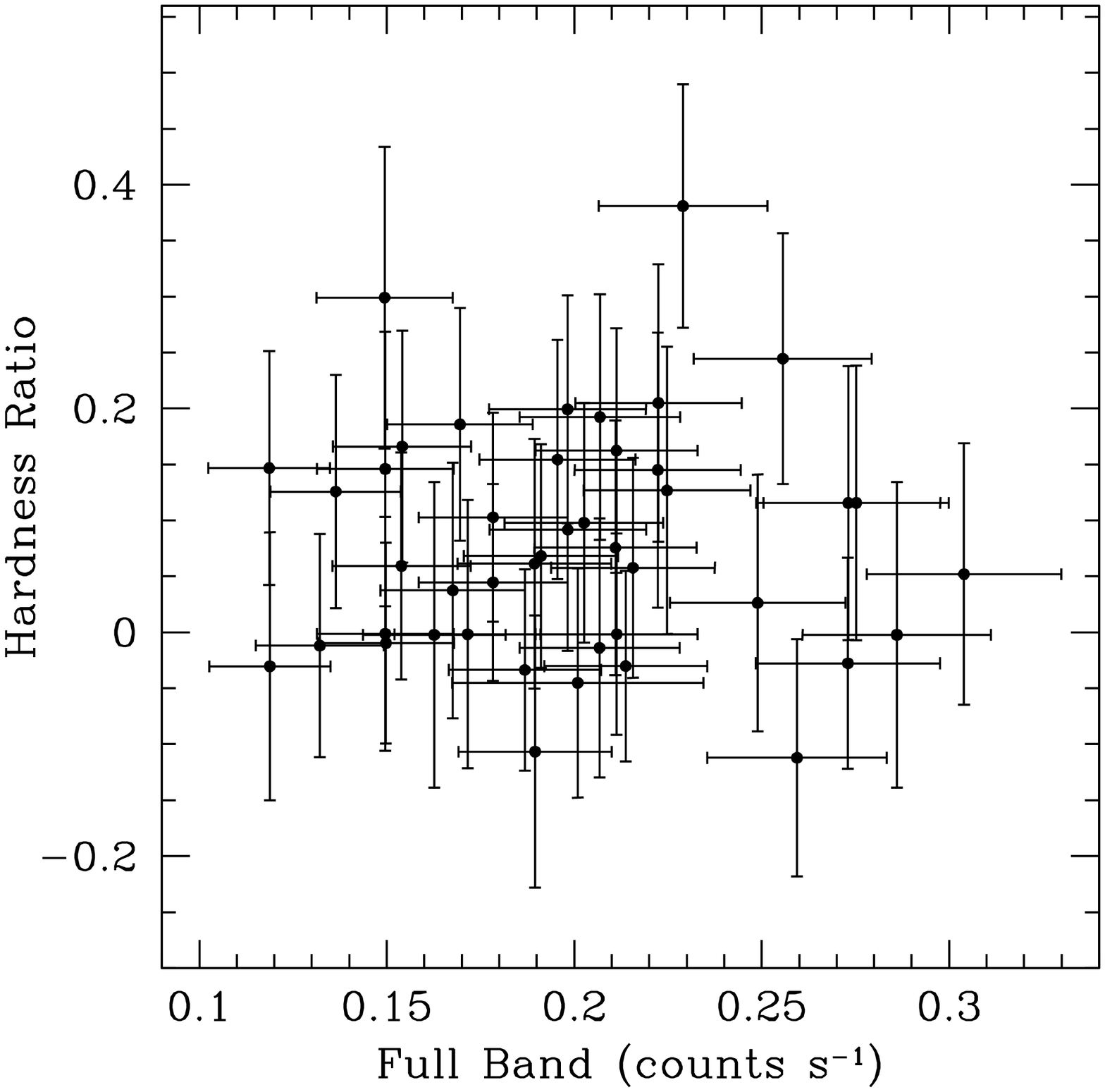}
\caption{{\it Left Panel:} Light curves of POX 52 taken from \chandra\ data. The top panel is 
the full (0.5--10.0 keV) light curve in counts s\per, the second panel is the soft 
(0.5--1.0 keV) X-ray light curve, the third panel is the hard (1.0--10.0 keV) X-ray light curve,
and the bottom panel is the hardness ratio (HR). {\it Right Panel:} Plot of hardness ratio 
versus the count rate in the full band.}
\label{ChandraLC}
\end{figure*}

\begin{figure*}
\begin{center}
\rotatebox{-90}{\scalebox{0.5}{\includegraphics{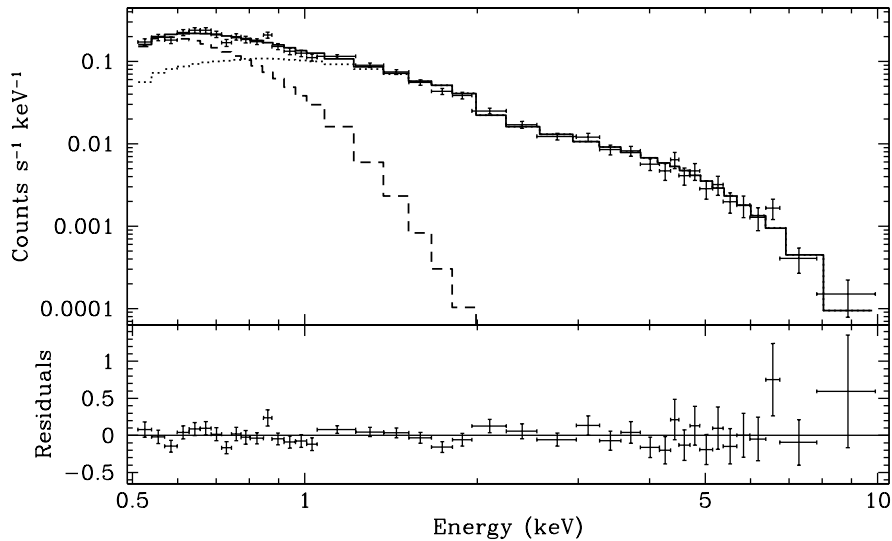}}}
\end{center}
\caption{The \chandra\ spectrum from 0.5--10.0 keV modeled ({\it thick line}) with an absorbed 
power law ({\it dotted line}) and disk blackbody ({\it dashed line}). Residuals are calculated
as Residual = (Data - Model)/Model.}
\label{ChandraX}
\end{figure*}

\begin{deluxetable*}{lcc}
\tablewidth{6in}
\tablecaption{X-ray Spectral Properties}
\tablehead{
\colhead{} & \colhead{\chandra} & \colhead{\xmmn}}
\startdata
$C$ statistic              & 183 (210 PHA bins, 205 dof) & 111 (104 PHA bins, 97 dof) \\
Photon Index $\Gamma$      & $1.8 \pm 0.15$              & $1.8 ^{+0.26}_{-0.24}$     \\
Disk $T_\mathrm{BB}$ (keV) & $0.13^{+0.03}_{-0.02}$      & $0.09^{+0.03}_{-0.01}$     \\
Foreground $N_H$  ($10^{21}$ cm\persq) & $0.50^{+1.1}_{-0.01} $  & $1.61^{+2.6}_{-1.2} $  \\
Partial Covering $N_H$  ($10^{21}$ cm\persq) &  \nodata & $58 ^{+8.4}_{-9.2}$        \\
Partial Covering Fraction (\%)   & \nodata   & $94^{+2.2}_{-3.3}$         \\
$F$(0.5--10 keV)  ($10^{-13}$ ergs cm\persq\ s\per)  & $12.6^{+0.05}_{-2.7}$    & $3.2^{+0.42}_{-1.51}$      \\
$L$(0.5--10 keV) ($10^{42}$ ergs s\per) & $1.50^{+0.01}_{-0.37}$    & $1.38^{+0.19}_{-0.82}$     
\enddata
\tablecomments{All model parameters were fit over a 0.5--10.0 keV energy range and we present 
observed flux and unabsorbed luminosity for both observations. For \xmmn, the spectra from both 
the EPIC-pn and MOS cameras were fit simultaneously.}
\label{xrayT}
\end{deluxetable*}

\section{X-Ray Observations}
\subsection{{\it Chandra}}
We first describe the \chandra\ observation since the source exhibited less spectral complexity
in the \chandra\ data than in the earlier \xmmn\ observation. X-ray observations of 
\dataset[ADS/Sa.CXO#obs/5736]{POX~52} were carried out using the Advanced CCD Imaging 
Spectrometer (ACIS) instrument on the {\it Chandra X-ray Observatory} for a total of 25 ks on 
2006 April 18 UT. The level 2 event file was restricted to the $0.5 - 10.0$ keV energy range, 
and there were no periods of high background found. POX~52 was observed to be consistent with a 
point source object, such that the FWHM of the PSF was $\sim 0\farcs5$, the 70\% and 90\%
encircled energy radii were $\sim 1.5$ and $2$ pixels, respectively\footnotemark, and the source
showed no extended emission (Figure \ref{ChandraPS}). 
\footnotetext{This is consistent with \chandra\ PSF analysis. For more information, see 
http://cxc.harvard.edu/proposer/POG/html/ACIS.html\#tth\_sEc6.6}
Source events were extracted using a circular region of radius 10 pixels ($4\farcs9$) centered
on POX 52. The source was too close to the edge of the CCD to use an annulus as the background 
region, therefore, a circular background region was extracted $46\arcsec$ away on the same chip 
with a radius of $24\farcs6$. There were approximately $1.2 \times 10^{-6}$ 
counts~s$^{-1}$~pixel$^{-2}$ within this background region, giving a background-subtracted 
average source count rate of $0.20$ counts s$^{-1}$. Due to the low source count rate, pile-up 
was not an issue. Data were analyzed using standard tasks, such as {\it dmextract} and 
{\it psextract} in CIAO (version 3.4/4.0$\beta$1; CALDB, version 3.4.0) and spectral models in 
XSPEC (version 12.3.0t). 

All light curves were created using time bins of 500 s and split into $0.5-10.0$, $0.5-1.0$, 
and $1.0-10.0$ keV energy ranges. These are denoted as the ``full,''  ``soft,'' and ``hard'' 
bands, respectively. The full-band light curve (Figure \ref{ChandraLC}, {\it top panel}) shows 
strong (factor of 2) variability over short ($\sim 500$ s) timescales. Also shown in Figure
\ref{ChandraLC} are the light curves for the soft band, hard band, and the hardness ratio, 
${\rm HR = (hard - soft)/(hard + soft)}$. There seems to be no significant correlation between 
hardness ratio and the flux from the full band as seen in the right panel of Figure 
\ref{ChandraLC} and is confirmed by the results of a Spearman rank correlation test. The 
correlation coefficient was $-0.16$, which describes a weak anti-correlation, with a 
significance of 0.27 that is essentially insignificant.  

The ACIS spectrum was first fit over the range $0.5 - 10.0$ keV with a simple absorbed power 
law, allowing all parameters to vary freely and optimized using the \citet{cash79} $C$ statistic
rather than $\chi^{2}$. The $C$ statistic is a better goodness-of-fit parameter when fitting low
signal-to-noise ratio spectra. The absorbed power law poorly fit the soft end of the observed 
spectrum, requiring another component to help model the excess of soft energies properly. We 
added a disk blackbody component to the model (\textsc{phabs*[powerlaw + diskbb]}), which 
resulted in an acceptable fit to the observed x-ray spectrum (Figure \ref{ChandraX}). The 
best-fitting model had a photon index of $\Gamma = 1.8 \pm 0.15$, and a disk blackbody 
temperature of $T_\mathrm{BB} = 0.13^{+0.03}_{-0.02}$ keV. The absorbing column density was 
$N_H = 0.50^{+1.1}_{-0.01} \times 10^{21}$ cm\persq, only slightly in excess of the predicted 
foreground column of $N_H \approx 0.4\times10^{21}$ cm\persq\ based on Galactic \ion{H}{1} maps
\citep{dl90,kal05}, as determined by the HEASARC online $N_H$ calculator\footnotemark.
\footnotetext{http://heasarc.gsfc.nasa.gov/cgi-bin/Tools/w3nh/w3nh.pl}

In place of the thermal disk blackbody, we tested fitting the spectrum with other thermal 
components plus the absorbed power law component, specifically the thermal bremsstrahlung model 
and the Raymond-Smith \citep{RS} plasma model. The bremsstrahlung model, with all parameters 
allowed to vary freely, produced a poor fit that strongly under-predicted the soft X-ray end of 
the spectrum by $\sim 3$ orders of magnitude. The Raymond-Smith model produced a reasonable fit 
to the data, similar to the fit when the disk blackbody was used, with a $C$ statistic of 182, 
nearly identical to the result for the combined absorbed power law and disk blackbody model 
($C = 183$). However, this model uses 4 free parameters (3 when the redshift is frozen) to fit 
the data compared to the 2 parameters used in disk blackbody model. Due to this difference, we 
select the combined absorbed power law and disk blackbody as our best-fit model for simplicity. 

There is possible evidence for an Fe K emission line seen in a single bin of energy 
${\rm E} = 6.6$~keV in the residuals of our best-fit model. We added a Gaussian component to our
model in order to determine constraints on the line energy and width. Allowing both the line 
energy and line width to vary freely, but fixing all other model parameters to our previous 
best-fit values, XSPEC modeled the line centroid at $6.71^{+ 0.07}_{- 0.12}$ keV with a line 
width ranging from $\sigma_{\rm line} = 0 - 0.82$ keV and an equivalent width (EW) of 
$0.3^{+ 0.9}_{- 0.3}$ keV. The large EW may be explained by the X-ray Baldwin effect (see Page 
\etal\ 2004, Jiang \etal\ 2007), but we cannot draw any firm conclusions, since the detection is
marginal, at best. \citet{Nandra07} analyzed \xmmn\ spectra of 26 Seyfert galaxies (mostly 
Seyfert 1) and compiled measurements of the Fe K$\alpha$ line. In this sample, the Fe K$\alpha$ 
energies ranged from 5.28 keV to 6.78 keV, with a range of equivalent widths of $0.009 -  0.394$
keV. Both the measured line energy and EW of the Fe K$\alpha$ from POX 52 are within the range 
of values measured by Nandra \etal, although the EW is at the high end of the values sampled. 
The large errors on these measurements make this a weak detection at best, and deeper 
observations are needed to obtain a definitive detection of the line.

\begin{figure*}
\epsscale{1.0}
\plottwo{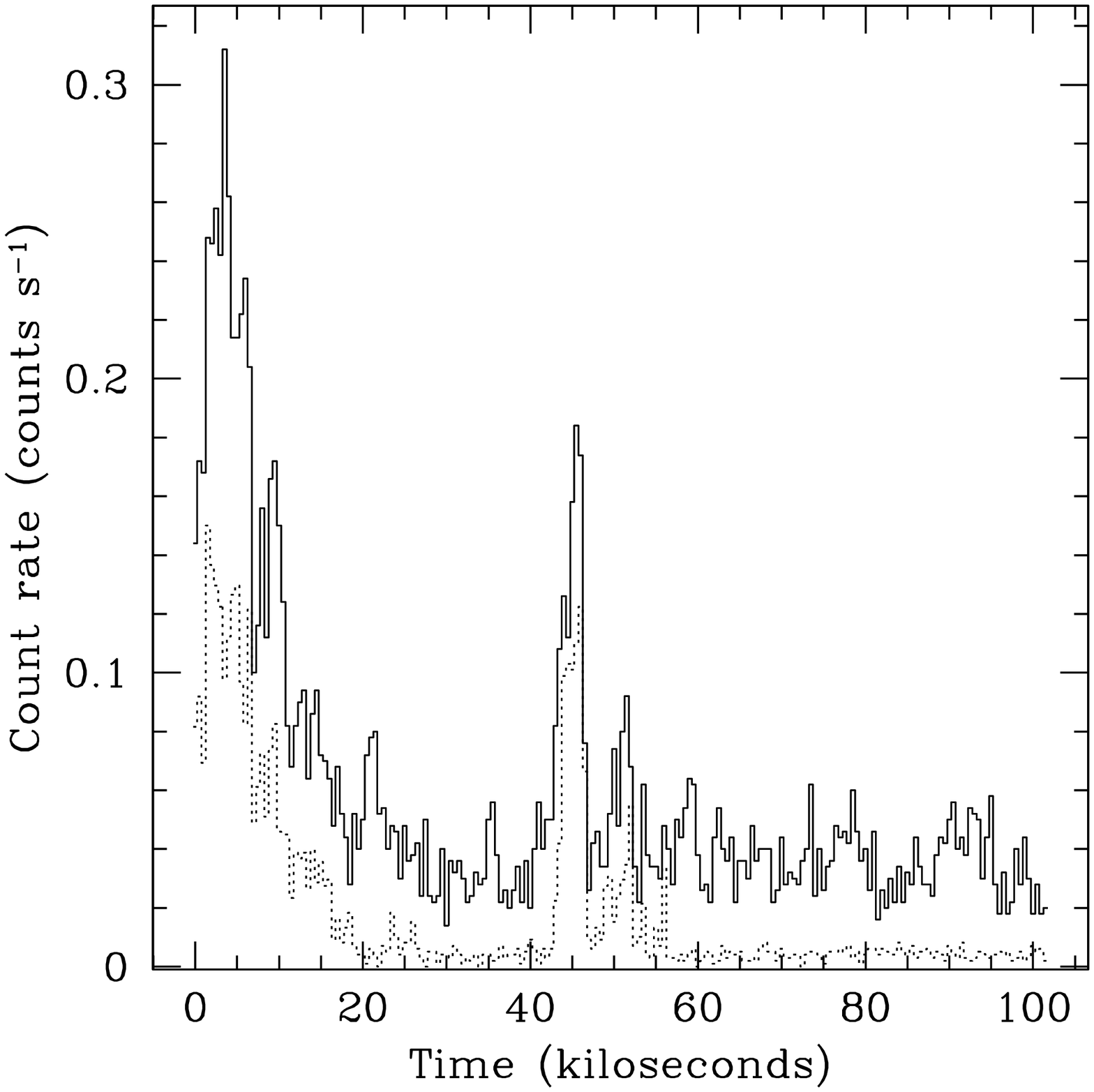}{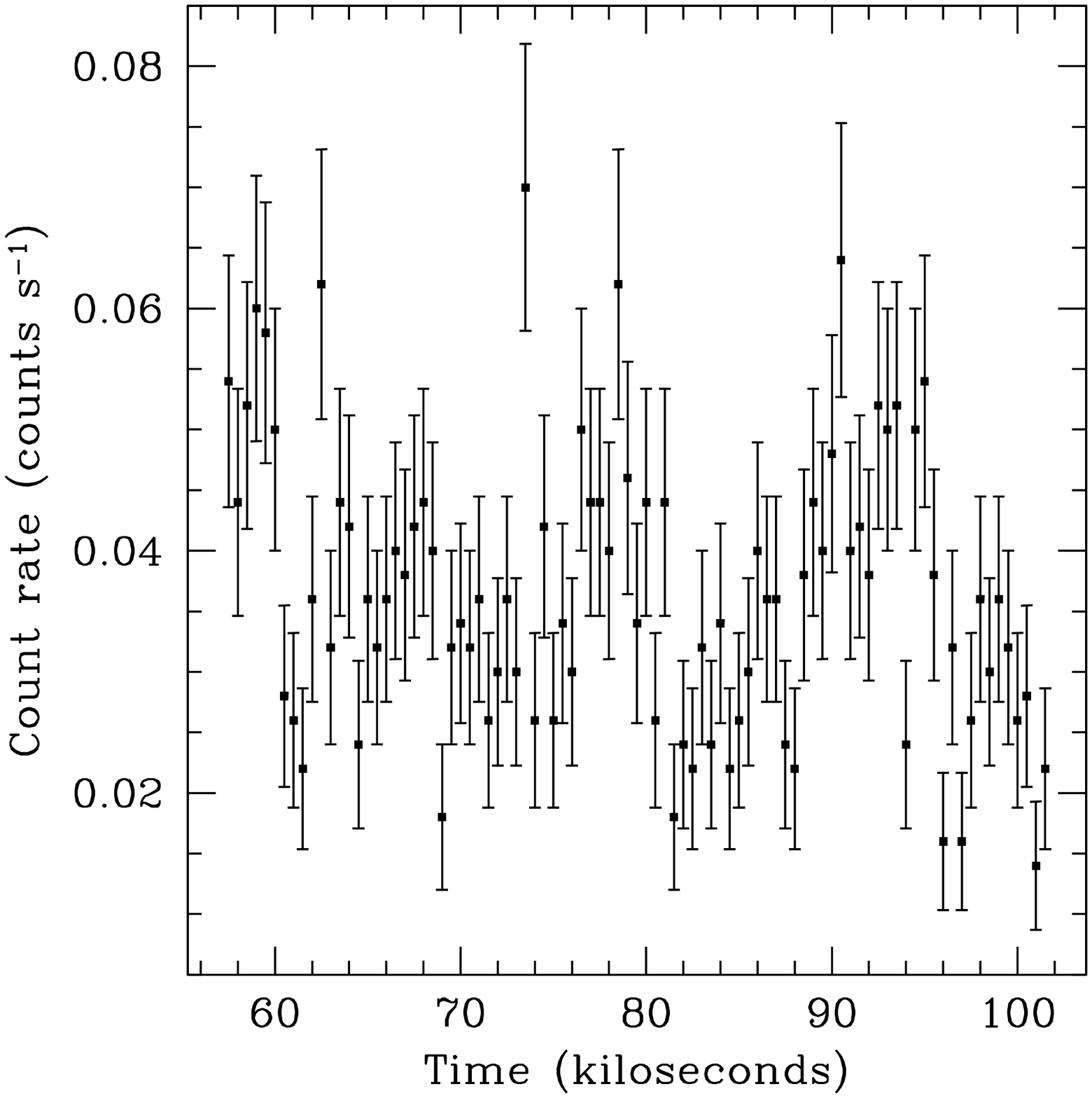}
\caption{{\it Left Panel:} The \xmmn\ EPIC-pn lightcurve (solid line) from 0.5--10.0 keV over 
the full 100 ks observation. The background spectrum is the dotted line. Note the large 
background flares at 0--10 ks and 40--55 ks.  {\it Right Panel:} The background-subtracted 
lightcurve of the second half of the observation starting at 57.5 ks, used in the spectral fit.}
\label{XMMLC}
\end{figure*}

\begin{figure*}
\begin{center}
\rotatebox{-90}{\scalebox{0.5}{\includegraphics{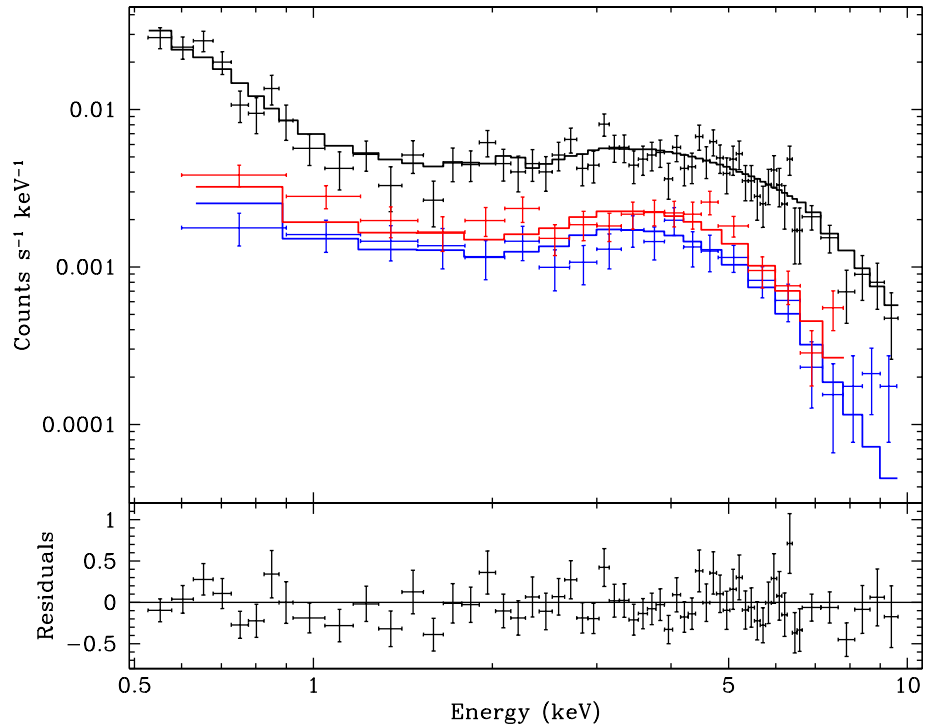}}}
\end{center}
\caption{\xmmn\ spectra over 0.5--10.0 keV, modeled with an absorbed power law, disk blackbody 
and a partial covering model with all parameters allowed to vary freely. The pn data is in 
black, MOS 1 is in blue, and MOS 2 is in red. The corresponding models are shown as solid lines 
in their respective colors. The bottom panel shows the residuals (calculated as Residual = 
[Data - Model]/Model) for the EPIC-pn data.}
\label{XMMX}
\end{figure*}

\subsection{{\it XMM-Newton}}
{\it XMM-Newton} observed POX 52 on 2005 July 8-9 UT for a total exposure time of 100 ks on each
of the 3 EPIC instruments: pn, MOS1, and MOS2. All data were reduced using the Science Analysis 
System (SAS, version 7.0.0) and XSPEC following the guidelines of the SAS Cookbook and SAS 
ABC-Guide. For all EPIC instruments, the source was extracted in a circular region with a radius
of $32\arcsec$, and due to the proximity of the source to the edge of the CCD in the EPIC-pn 
camera, the background was extracted from a circular region free of sources, located 
$107\arcsec$ away from the source with a radius of $44\farcs9$. Only events corresponding to 
patterns $0 - 4$ (single or double pixel events) were used for the pn event file and patterns 
$0~-~12$ (single, double, triple and quadruple pixel events) for the MOS event files. All event 
files were restricted to an energy range of $0.5 - 10.0$ keV to better compare with the 
\chandra\ data. For all three instruments, events were excluded that occurred next to the edges 
of the CCD or next to bad pixels. Again, the source count rate was low enough to neglect effects
due to pile-up.

Again using 500 s time bins, a light curve (Figure \ref{XMMLC}) was created using the 
entirety of the pn dataset. However, the first half of the observation contained two large 
background flares and due to this contamination, only the second half of the observation 
(Figure \ref{XMMLC}, {\it right panel}) was used in the spectral analysis (where the total 
counts in the background region are $\sim 11.5\%$ of the source counts). Examination of the 
second half of the observation shows a similar amplitude of variability to the \chandra\ 
observations, but over slightly longer ($10^3 - 10^4$ s) timescales. 

Observations of POX 52 with \xmmn\ yielded a significantly lower flux than the subsequent 
\chandra\ data. Figure \ref{XMMX} shows the spectra from both the pn and MOS cameras. There is a
hint of a detection of a narrow Fe K emission line near 6.4 keV (contained in the bin centered 
at 6.375 keV with a width of 0.125 keV), as can be seen in the EPIC-pn residuals 
(Figure \ref{XMMX}, {\it lower panel}). A simple absorbed power law and a combined absorbed 
power law plus disk blackbody both proved to be poor fits, even when the same parameter values 
were used from the Chandra fit, under-predicting the spectral shape by $1 - 2$ orders of 
magnitude below $\sim 1$ keV. 

The \xmmn\ spectrum is harder than in the \chandra\ observation and requires more absorption to 
fit the spectral shape. Adding a partial covering component to the model 
(\textsc{phabs*pcfabs*[powerlaw + diskbb]}) improved the fit significantly. We fit the data 
using this model, allowing all parameters to vary freely. The photon index and blackbody 
temperature of this model are consistent within the errors with values derived from the 
\chandra\ data. The major difference is the need for much higher levels of absorption. The 
absorbing column density is three times that needed to model the \chandra\ data and the model 
adds a partial covering absorbing column density of 
$58^{+8.4}_{-9.2} \times 10^{21}~{\rm cm}^{-2}$ with a partial covering fraction of 
$94^{+2.2}_{-3.3}\%$. Since the partial covering fraction is very high, we tested forcing the 
fraction to 100\% or simply using one larger absorbing column density, but both scenarios tended
to over-absorb the power law in the soft X-ray end. The unabsorbed luminosity calculated for the
\xmmn\ observation is within the 1$\sigma$ uncertainty of the \chandra\ unabsorbed luminosity, 
which suggests that the properties of the central engine itself did not change significantly 
between these two observations. The factor of 4 flux difference between the two X-ray 
observations is then explained by the additional absorption needed to fit the \xmmn\ spectra.

\begin{figure*}
\begin{center}
\scalebox{1.0}{\includegraphics{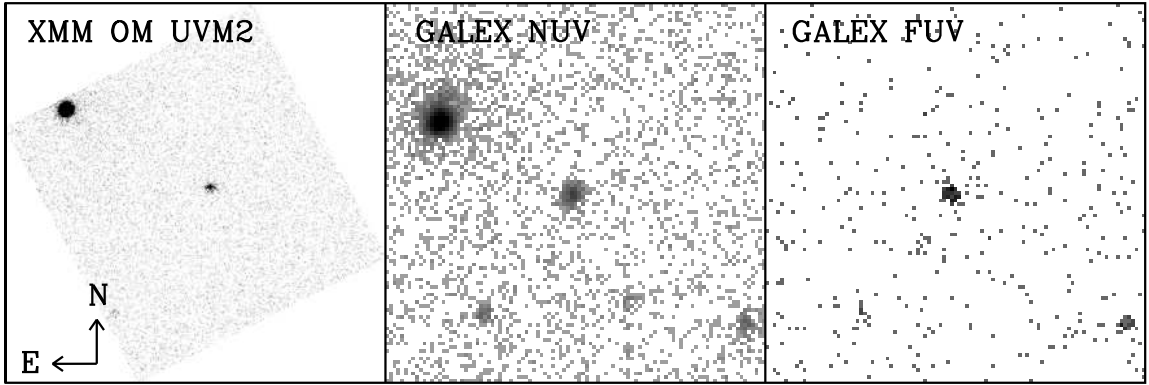}}
\end{center}
\caption{Ultraviolet images of POX 52 (central object in each panel). Each image is 150\arcsec\ 
on a side. The OM panel is a single 1 ks image taken from our sample of 72 images.}
\label{UVimage}
\end{figure*}

\begin{figure}
\epsscale{1.0}
\plotone{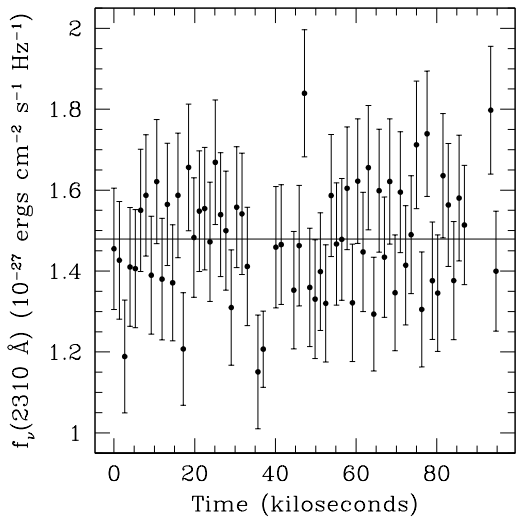}
\caption{UV light curve of POX 52 from the Optical Monitor onboard \xmmn. The horizontal line
represents the mean observed flux.}
\label{OMLC}
\end{figure}

\section{Ultraviolet Observations}
\subsection{\xmmn\ Optical Monitor}
The Optical Monitor (OM) on \xmmn\ took a series of 72 images simultaneously with the X-ray 
observation, using the UVM2 filter, which is centered at 2310 \AA, extending from 
$\sim 1900-2950$ \AA\ (FWHM of the PSF $= 1\farcs8 \sim 4$ pixels). Each of these images had 
an exposure time of 1 ks. The images were processed using the \xmmn\ SAS routine {\it omichain} 
to output calibrated photometric data via aperture photometry. Out of 72 images, POX~52 was 
easily detected in 60 images by the source detection algorithm in {\it omichain} and can be seen
as the central object in Figure \ref{UVimage}. Five additional images were reprocessed with 
{\it omprep, omdetect, ommag, ommodmap,} and {\it ommat} separately in order detect POX~52 by 
varying the detection threshold. For the remaining 7 images, the \emph{omichain} routine was 
unable to detect or measure the source count rate due to high background count rates, and these 
images were discarded.

We used the photometric results of {\it omichain} to construct a UV light curve, shown in Figure
\ref{OMLC}. The measured flux was corrected for Galactic extinction by using the 
\citet{Schlegel} reddening maps and averaging the \citet{CCM} extinction law over the OM 
passband resulting in $A_{\rm OM} = 0.32$ mag for the UVM2 filter. The mean flux density over 
the set of observations is $f_\nu({\rm 2310}$~\AA$) = (1.5 \pm 0.1) \times 10^{-27}$~\ergcmsH. 
Inspection of the light curve shows that the source is not highly variable in the UV. Only 25\% 
of the individual points differ from the mean flux level by more than $1\sigma$, and only 3/64 
(or 5\%) of the points differ from the mean flux by more than $2\sigma$; there are no $3\sigma$ 
outliers. The observed variability is therefore generally consistent with the expected level of 
variations from photon-counting statistics for a constant source. To further quantify any 
possible variability, we calculated the normalized excess variance of the light curve, following
the prescription described by \citet{Nandra}:
\begin{equation}
\sigma^2_\mathrm{nxs} = \frac{1}{N \mu^2} \sum_{i=1}^{N} [(X_i - \mu)^2 - \sigma_i^2],
\label{excessvariance}
\end{equation}
where $X_i$ denotes the count rate of the $i$-th point in the light curve, $\sigma_i$ is its 
uncertainty, $\mu$ is the mean of the $X_i$ values over the entire light curve, and $N$ is the 
number of points in the light curve. The resulting excess variance is very small and actually 
has a negative value: $\sigma^{2}_\mathrm{nxs} = -0.0029$. This result confirms that there is no
evidence for significant variability in the UV light curve. The negative value of 
$\sigma^{2}_{\rm nxs}$ implies the error bars on the measurements are most likely slightly 
overestimated by the SAS processing and detection algorithm.

\subsection{{\it GALEX}}
As part of the \emph{GALEX} All-sky Imaging Survey (AIS), POX~52 was imaged simultaneously in 
the near-UV (NUV) and far-UV (FUV) bands on 2006 February 9 UT for a total of 114 s. The NUV 
and FUV bands span the ranges $1771 - 2831$ \AA\ and $1344 - 1786$ \AA, respectively, with 
effective wavelengths of 2271 \AA\ and 1528 \AA. The FWHM of the PSF for the NUV filter is 
$4\farcs0$ ($\sim 2 - 3$ pixels) and for the FUV filter is $5\farcs6$ ($\sim 3 - 4$ pixels). 
Figure \ref{UVimage} shows both NUV and FUV images of POX 52 alongside one taken with the OM on 
\xmmn. Galactic extinction was calculated using the same method as for the OM, resulting in 
$A_{\rm NUV} = 0.43$ mag and $A_{\rm FUV} = 0.84$ mag. These were then used to correct the 
fluxes of POX 52 taken from the AIS catalog, resulting in a NUV flux density of 
$f_{\nu}(2271$~\AA$) = (1.88 \pm 0.08) \times 10^{-27}$ \ergcmsH\ and a FUV flux density of 
$f_{\nu}(1528$~\AA$) = (1.82 \pm 0.10) \times 10^{-27}$ \ergcmsH. The NUV band of {\it GALEX} 
surrounds the UVM2 filter on the OM, giving us the opportunity to check for long-term 
variability in the UV. The flux densities calculated from both instruments are consistent within
the $2\sigma$ uncertainty range, meaning we see no significant change in UV luminosity in the 7 
months between the separate observations. 

In order to estimate the UV host galaxy contribution, we take the observed range of NUV$-V$ and 
FUV$-V$ colors of Virgo dwarf elliptical galaxies from \citet{Boselli} and compute possible NUV 
and FUV flux densities based on the previous measurement of $V_{\rm host} = 17.4$ mag for POX~52
from \citet{bar04}. This results in a estimated host galaxy magnitude in the range of 
NUV $= 20.4 - 22.4$ mag (AB) and FUV $= 21.9 - 24.9$ mag (AB), which we convert to a host galaxy
NUV flux density of $f_{\nu}(2271$~\AA$) = (2.5 - 4.0) \times 10^{-29}$ \ergcmsH\ and a FUV flux
density of $f_{\nu}(1528$~\AA$) = (4.0 - 63) \times 10^{-30}$ \ergcmsH\ for POX~52. We take the 
average flux density value in each of these bands to correct for host galaxy contamination and 
find corrected AGN flux densities of $f_{\nu}(2271$~\AA$) = 1.85 \times 10^{-27}$ \ergcmsH\ and 
$f_{\nu}(1528$~\AA$) = 1.80 \times 10^{-27}$ \ergcmsH\ for the NUV and FUV, respectively. 
Correcting the OM measurement in a similar manner results in a negligible change to the flux 
density. We note that, even for the highest estimate of host galaxy UV flux, the source is still
very AGN dominated in the UV.

\section{Radio Observations}
POX 52 was observed for one hour on 2004 October 10 UT with the VLA in A configuration 
\citep{Thompson}, as part of a larger program to observe low-mass active galaxies from Greene \&
Ho (2004; see Greene \etal\ 2006). Observations were carried out at 4.860 GHz (6 cm; C band) 
with a bandwidth of 50 MHz for each of two intermediate frequencies separated by 50 MHz. For
purposes of phase calibration, observations of the target were alternated every 3-5 minutes with
observations of a bright radio source (PKS J1159-2148, located $\sim 1\arcdeg$ away). Our 
overall flux scale is tied to 3C 48, with an assumed flux density of 5.4 Jy, and uncertainties 
in the absolute fluxes of the phase calibrators, estimated to be $\sim 5\%$, dominate the flux 
scale uncertainties.

After removal of corrupted records, flux and phase calibration were performed within AIPS 
\citep{Greisen}.  We then Fourier transformed the observed visibilities to create an image of 
intensity on the sky, using a pixel scale of $0\farcs06$ and a resulting synthesized beam of 
$\Delta \theta \approx 0\farcs6$.  We find no confusing sources within the primary beam 
($\sim 9$\arcmin), and so, because deconvolution algorithms such as CLEAN may change the noise 
statistics in a source-free image, we have calculated the upper limit of $0.078$ mJy directly 
from the direct Fourier transform. This upper limit represents $3 \times$ the measured noise 
limit in the uncleaned map of $= 0.026$ mJy and corresponds to a radio power of 
$P < 8.1 \times 10^{19}\ {\rm W\ Hz^{-1}}$ and a radio-loudness parameter 
($R \equiv f_{6 {\rm cm}}/f_{4400 {\rm \AA}}$) for POX~52 of $R < 0.28$, where a radio-loud 
object is usually defined to have $R \geq 10$ \citep{K89}.

These values are consistent with the majority of the other GH sample sources studied by 
\citet{GHU}. Out of 19 objects studied, Greene \etal\ only significantly detected one object, 
with $3\sigma$ upper limits to the radio power of the other 18 objects around 
$P < {\rm a~few}~\times 10^{21}~{\rm W\ Hz^{-1}}$. The high upper limits for the GH objects 
relative to POX 52 are due to the larger distances to the GH sample. The upper limits on the 
radio-loudness parameter may be a better comparison between POX 52 and the GH objects, which 
range from $R < 0.68$ to $R < 9.9$ for the GH sample, but the majority of these objects are
well below the $R \geq 10$ cutoff that usually describes a radio-loud object. Stacking the GH 
sample results in $R < 0.27$, giving a stricter upper limit to the radio-loudness of the sample 
as a whole.

\section{Infrared Observations}
\subsection{\it Spitzer Space Telescope}
\begin{figure}
\begin{center}
\rotatebox{-90}{\scalebox{0.3}{\includegraphics{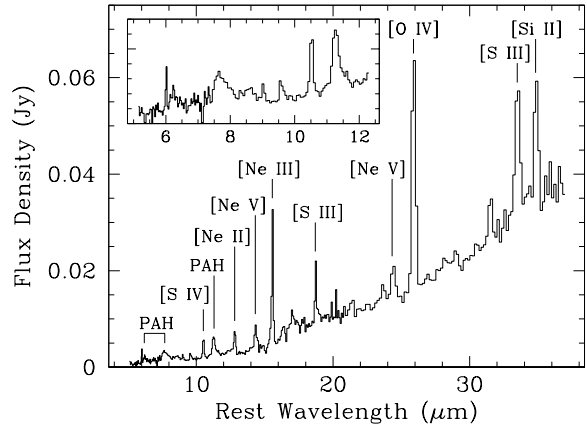}}}
\end{center}
\caption{{\it Spitzer} IRS spectrum of POX 52 from $5-38$ \um. The most prominent spectral lines
are identified. The inset shows a close-up of the $5-12$ \um\ region.}
\label{Spitzer}
\end{figure}

POX 52 was observed on 2006 July 24 UT with the {\it Spitzer} Infrared Spectrograph (IRS) 
\citep{IRS} as part of a sample of low-mass Seyfert galaxies (Program ID: 30119). The analysis 
of the entire sample, including measurements of emission lines and continuum shape, will be 
presented in a later paper, but we include the POX 52 spectrum here in order to supplement our 
multiwavelength coverage of this object. It was observed in the Short-Low and Long-Low slit 
modes, which correspond to a total wavelength range of $5.2 - 38$ \um.  Exposure times were 
$2 \times 60$ s each in the SL1 and SL2 settings, and $2 \times 120$ s each in the LL1 and LL2 
settings. Data were processed prior to download with IRS Pipeline Processing version 15.3.0. 
Data were first cleaned using the IDL routine IRSCLEAN (v.\ 1.9), and the images were then 
coadded following the procedure in the {\it Infrared Spectrograph Data Handbook}\footnotemark\ 
(v.~3.1). The spectrum from each slit was extracted and calibrated using SPICE (v.~2.1.2). For 
each slit, the two nod positions were subtracted from each other in order to sky subtract the 
resulting data. Spectra from each of the four slits were combined and any overlapping data from 
adjacent wavebands were averaged together in the final product. Data from the LL slit was 
multiplied by a factor of 1.4 in order to properly match data from the SL slit. Figure 
\ref{Spitzer} shows the combined spectrum with the most prominent emission lines identified, 
including [\ion{Ne}{2}] $\lambda 12.81$~\um, [\ion{Ne}{3}] $\lambda 15.56$~\um, [\ion{Ne}{5}] 
$\lambda 14.32$~\um, [\ion{O}{4}] $\lambda 25.89$~\um, [\ion{S}{3}] 
$\lambda\lambda 18.71, 33.48$~\um, [\ion{S}{4}] $\lambda 10.51$~\um, [\ion{Si}{2}] 
$\lambda 34.82$~\um, and PAH features at 6.2 \um, 7.7 \um, and 11.3 \um.   

The IRS spectra of POX 52 shows possible weak silicate absorption near $10$ \um\ seen in some
Seyfert 1 galaxies, but more commonly associated with Seyfert 2 galaxies \citep{Hao}. Assuming a
power law fit over the $30 - 20$ \um\ range with $f_{\nu} \propto \nu^{-\alpha}$, POX 52 shows a
steeper spectral slope $\alpha_{30-20} \approx 2.1$ than many Seyfert 1 galaxies 
\citep{Buchanan}. The spectral slope from $15-6$ \um\ of $\alpha_{15-6} \approx 1.16$ is 
significantly flatter than $\alpha_{30-20}$ suggesting the best continuum fit to be a broken 
power law, which is often seen in Seyfert 1 spectra \citep{Weedman}. The presence of PAH 
features and the observed red continuum (seen as the steep $\alpha_{30-20}$ slope) suggests 
possible star formation in the host galaxy, however, we defer any further analysis to the later 
paper on the full {\it Spitzer} sample. 
\footnotetext{http://ssc.spitzer.caltech.edu/irs/dh/}

\subsection{2MASS Point Source Catalog}
Near-infrared magnitudes were taken from the 2MASS Point Source Catalog \citep{2MASS} in the $J$
(15.7 mag), $H$ (15.0 mag), and $K_S$ (14.5 mag) bands to be included along with our other data 
in the SED (Figure \ref{SED}). These magnitudes include flux from the 
host galaxy as well as the AGN. In order to correct for this we used averaged  
$B - J = 2.05,\ B - H = 2.57,$ and $B - K = 2.67$ color indices derived from photometry of 10 
dwarf ellipticals in the Virgo Cluster \citep{james94}. Our GALFIT estimate of the $B$-band 
$m_{\rm host}$ (corrected from the F435W ACS/HRC filter to the standard Johnson $B$ filter using
the IRAF routine {\it calcphot}) was used in order to estimate the $JHK$ contribution from the
host galaxy. The host galaxy contributions were then subtracted from the 2MASS magnitudes, 
resulting in the following AGN point source $JHK$ magnitude estimates:
$J_{\rm AGN} = 16.8^{+0.8}_{-0.6},\ H_{\rm AGN} = 15.8^{+0.8}_{-0.5},\ {\rm and}\ K_{\rm 
AGN} = 15.0^{+0.6}_{-0.8}$ mag. The errors were calculated based on the range of 
$B - J,\ B - H,$ and $B - K$ colors for dwarf elliptical galaxies derived from \citet{james94}.

\section{Discussion}
\subsection{Host Galaxy Structure}
\citet{bar04} previously classified POX 52 as a dwarf elliptical based on its morphology, their 
measurements of $M_B = -16.8$, $r_e = 1\farcs2$, and $n=3.9$, and the location of POX~52 in 
the ``$\kappa$-space'' projections \citep{BBF, Burstein97} of the fundamental plane. These 
$\kappa$ parameters combine stellar velocity dispersion, surface brightness, and effective 
radius together in such a way, which correspond to the physical parameters of stellar mass and 
mass-to-light ratio, that a clear split can be seen among giant ellipticals, dwarf ellipticals, 
dwarf spheroidals, and globular clusters. In these projections, POX~52 falls nearest the Virgo 
dwarf elliptical galaxies from \citet{Geha03}, and is offset only slightly toward a higher 
surface brightness and a lower mass-to-light ratio. Our new \hst\ observations do little to 
change the location of POX 52 in these projections, as the surface brightness and effective 
radius found in this paper are consistent with the previous values. 

The addition of similar objects from the GH sample, based on new \hst\ imaging \citep{GH08}, to 
these fundamental plane relations suggests that they are more centrally concentrated than 
typical dwarf ellipticals. However, in plots of $\mu_e$ vs.\ $M_V$ and $r_e$ vs.\ $M_V$, they 
reside in between the distributions of dwarf ellipticals and classical ellipticals, as a 
possibly intermediate population. These galaxies clearly have properties that are in common with
both populations, but they follow the distribution of dwarf ellipticals when kinematic data are 
combined with photometric data in the ``$\kappa$-space'' projections and therefore, we conclude
that POX 52 and similar objects are most closely related to dwarf ellipticals, although they 
remain unusual for having unexpectedly highly \sersic\ indices.

We note a difference in terminology between what some authors call a dwarf elliptical galaxy 
\citep{BJ, GGM} and a different designation, spheroidals, in order to distinguish the typically 
less centrally concentrated objects from low-luminosity versions of classical ellipticals 
\citep{Kor85}. In this paper, we refer to these galaxies as dwarf ellipticals for consistency 
with the nomenclature adopted in our earlier paper \citep{bar04}.

From our GALFIT decompositions, we can conclude that the host galaxy is well fit in both of the
observed filters by a \sersic\ profile with $n \approx 4$ and $r_e \approx 1\arcsec$ at 
$r > 0\farcs5$. The structure of the inner region of the host galaxy remains complicated due to 
the second \sersic\ component needed to provide a reasonable fit in the F435W filter. We noted 
earlier that the inner \sersic\ component seems to be compensating mostly for the PSF mismatch, 
but its broader profile suggests the possibility of a spatially compact, blue component, such as
a nuclear star cluster. Looking at the $B-I$ color of POX~52 with and without the inner \sersic\
component included, we can bracket the range of possible colors of the host galaxy. Excluding 
the inner \sersic\ component, we find $B-I = 2.0$ for the host galaxy, which is similar to 
fiducial values ($B-I = 2.0-2.1$) for S0/Sab galaxies found in \citet{Fukugita} and from 
{\it synphot} calculations ($B-I = 2.2-2.3$) with the \citet{KC} S0/Sab galaxy templates. If we 
assume that the compact component is entirely due to starlight, the resulting color of 
$B-I = 0.74$ is bluer than both the \citet{Fukugita} fiducial value for Im galaxies 
($B-I = 0.9$)and all of the starburst galaxy templates ($B-I = 0.9 - 1.6$) from \citet{KC}. 
Therefore, we conclude that the inner \sersic\ component in the F435W image must contain a 
substantial amount of light from the AGN point source. However, we cannot determine the exact 
fractions of AGN and starlight in this component. This leaves a wide range of possible colors 
for the host galaxy.

\subsection{X-ray Spectral Properties}
X-ray spectral variability is not an unusual occurrence in an AGN, but is normally manifested
in small changes in the spectral slope or in the width or amplitude of an emission line. 
Strong variability in the absorbing column density is more rare, but is often attributed to 
partial covering. In this scenario, there exists a patchy absorber surrounding the AGN made 
up of many smaller gas clouds orbiting the central engine with close to Keplerian velocities.
The variations seen in column density are therefore related to the number of clouds along the 
line of sight. This has become a common model to use when a change in the spectral slope on the 
soft end is observed, as has been seen in several NLS1 galaxies, including IRAS~13224-3809 
\citep{Boller03, GB04}, RX~J2217.9-5941 \citep{Grupe04}, Mrk~335 \citep{Grupe07}, Mrk~1239 
\citep{GMK04}, and 1H~0707-495 \citep{Gallo04}. 

The bright NLS1 Mrk~335 has recently experienced a change in absorption very similar to what we 
observe in POX~52. Mrk~335 has regularly been studied in the X-rays since 1971 with no dramatic 
changes in flux until 2007, when \citet{Grupe07} measured a factor of 30 decrease in flux 
compared to all previous observations. When the spectra were modeled, Grupe \etal\ were unable 
to obtain a good fit with scaled-down versions of previously used models and found that adding a
partial covering model to an absorbed power law with a high partial covering fraction of 
$0.9 - 1$, much like that of POX 52, enabled an acceptable fit. When corrected for this 
additional absorption, the change in intrinsic luminosity is only a factor of $4 - 6$ when 
compared to earlier observations. Grupe \etal\ note that a partial covering model can also 
reasonably fit their 2006 data with a lower partial covering fraction of $\sim 0.5$, but does 
not result in the best fit for their 2000 data. This suggests the partial covering absorber was 
moving into the line of sight as early as 2006.

Assuming the sizes of these discrete absorbing clouds do not vary much from one source to 
another, and that the accretion disks and surrounding X-ray-emitting regions of less massive 
black holes like POX 52 are, in fact, scaled-down versions of those associated with their more 
massive counterparts, one would expect the X-ray absorbing region to be less voluminous and 
therefore contain fewer absorbing clouds. Furthermore, if the differences in column density are 
attributed to the number of clouds instead of the size of clouds, it is possible that an 
X-ray-emitting region surrounding a black hole the size of POX 52 would have a higher likelihood
of being observed with almost no absorbing clouds in the line of sight than for a more massive 
source due to the differences in the number of clouds associated with each source. The black 
hole in POX 52 is $\sim 2$ orders of magnitude less massive than that in Mrk 335 
\citep{Pet98, Pet04, V} and moved from an unabsorbed state to the partial covering state in less
than 9 months, whereas the change in absorbing column in Mrk~335 occurred over $\sim 1 - 7$ 
years \citep{Grupe07}. These observations give upper limits to the timescales of absorption 
variability based on the partial covering scenario. 

\citet{bar04} examined the {\it ROSAT} all-sky survey image of the position of POX~52 with an 
exposure time of $\sim 90$ s. Although there was no prominent source at the position of POX~52, 
Barth \etal\ showed there were two counts above the background rate within a circular aperture 
of radius equal to $\sim$ FWHM of the PSF of the detector, for a $3.0 - 3.5 \sigma$ detection. 
With this in mind, we used the WebPIMMS\footnotemark\ online calculator to determine what state 
POX 52 might have been in when this observation was taken. Inputting our best-fit models from 
\xmmn\ and \chandra, we find that the unabsorbed state would have resulted in a $0.4 - 2.4$ keV 
{\it ROSAT} detection in the range of $4 - 14$ counts, and the absorbed state would have 
produced $1 - 4$ counts during the observation. From this we conclude that the ROSAT flux is 
consistent with the \xmmn\ absorbed state, or possibly a lower luminosity state. Further 
observations would help determine the more common X-ray state of POX~52 and better constrain the
timescales of this variability. 
\footnotetext{http://heasarc.nasa.gov/Tools/w3pimms.html}

We see no change in the intrinsic power law spectral slope of $\Gamma \approx 1.9$ between these
two observations. This is somewhat unusual given the large amount of both short-term 
($\sim 500$ s) and long-term ($\sim 10^7$ s) variability that characterizes many NLS1 galaxies, 
including POX 52. For example, POX 52's most similar companion, NGC 4395 has been shown to have 
strong variations not only in flux (factor of 2 over 5 months, O'Neill \etal\ 2006), but also in
spectral slope ($\Gamma \approx 0.6-1.7$ over 2 years, Moran \etal\ 2005). A $2-10$ keV spectral
slope of 0.6 is unusually low for either a Seyfert 1 galaxy or a NLS1 galaxy; a slope of 
$\sim 2$ is far more common for Seyfert 1, NLS1 galaxies, and quasars \citep{Crummy, Lei, Pic}. 
The temperature of the blackbody component, used to model the soft excess seen in many NLS1 
galaxies, also remained unchanged for POX 52 between the two observations, with a temperature of
$kT = 0.13$ keV that is within the normal range ($0.1 - 0.3$ keV) seen in NLS1 galaxies 
\citep{Lei}. When NLS1 spectra with soft excesses are fit with a model consisting of only a 
power law or absorbed power law component, the result is a steep spectral slope of 
$\Gamma \approx 3-5$ \citep{Boller96}, as opposed to the more common NLS1 spectral slope of 
$\Gamma \approx 2$. 

In comparison, the similar low-mass GH sample objects observed with \chandra\ also have 
slightly different X-ray properties \citep{GH07a}. These 10 objects were observed for 5 ks 
each, resulting in only 5 objects with enough counts to allow for spectral fitting, and even 
this could only be done in the $0.3 - 5$ keV range. A soft photon index was derived from the 
hardness ratios for each object, giving a range of $1 < \Gamma_S < 2.7$ for the sample which 
brackets the $0.5 - 10.0$ keV photon index for POX 52. The GH sample objects show no signs of 
either a soft excess or intrinsic absorption. The observations did not allow for study of 
short-term variability due to low count rates. 

\citet{Mini} re-observed four of these GH objects with \xmmn\ for 20 - 30 ks, in order to obtain
a high enough signal-to-noise ratio for spectral fitting. These objects have remarkably similar 
spectra in the $0.5 - 10.0$ keV range to the \chandra\ spectrum of POX~52, and fits with power 
law plus disk blackbody models produce photon indices in the range of $1.8 < \Gamma < 2.5$ and
temperatures $\sim 0.11$ keV. Although the blackbody temperature of the soft excess is similar 
to what is seen in both the \chandra\ and \xmmn\ observations of POX~52, the hard photon index 
of these GH objects is, on average, slightly harder than what has been observed in POX~52. These
objects are an order of magnitude more luminous in the X-ray than POX~52, and contain black 
holes with masses around an order or magnitude larger than POX~52 \citep{GH07c}.

\begin{figure}
\epsscale{1.0}
\plotone{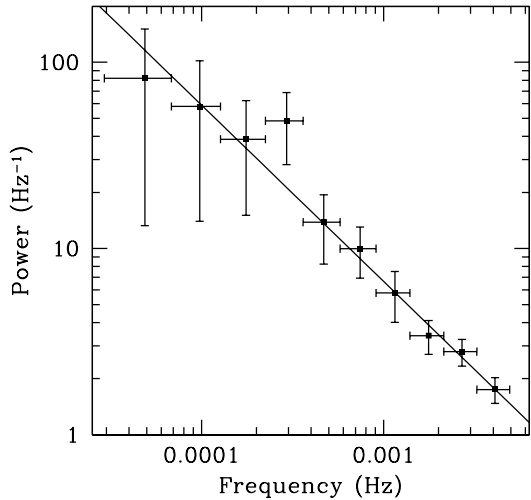}
\caption{PSD derived from \chandra\ data using a light curve with 100 s time bins. The PSD 
itself has been binned by a factor of 5. A power law with slope $-0.95$ is shown as a solid 
line.}
\label{PSD}
\end{figure}

\subsubsection{Power Spectral Density}
AGNs are highly variable in X-rays and with good data sampling, one can study the frequency 
dependence of these variations by computing the power spectrum density (PSD), which is a measure
of the amount of variability at a given frequency. The PSDs of AGNs and Seyferts are often fit 
with a combination of two power laws with a slope of $-2$ at the high-frequency end and then 
breaking to a slope of $-1$ at the low-frequency end. The frequency at which this change occurs 
is known as the break frequency, $\nu_{B}$, and appears to correlate inversely with black 
hole mass \citep{Mark}. 

Since the \chandra\ data have a higher signal-to-noise ratio than the \xmmn\ data and more 
closely represent the intrinsic state of the central engine without the heavy obscuration seen 
with \xmmn, we chose to carry out the PSD analysis on the \chandra\ dataset. We start by 
creating a light curve using 100 s time bins covering the $0.5 - 10.0$ keV energy range. The 
Xronos (v5.21) tool {\it powspec} (v1.0) was used to create the PSD from the light curve and can
be seen in Figure \ref{PSD} binned by a factor of 5 in frequency in order to increase the 
signal-to-noise ratio. There may be a possible flattening of the PSD from low frequencies and a 
break to a steeper slope at $\sim 3 \times 10^{-4}$ Hz. However, the PSD is easily fit with a 
single power law with a best-fit slope of $-0.94$ and a $\chi^{2} = 2.87$ for 8 dof. This fit is
good enough that a broken power law fit is not justified. To test whether any constraints on 
$\nu_{B}$ could be derived, we attempted to fit a broken power law to the data. Unfortunately, 
all attempts at fitting the data while allowing parameters to vary freely produced either a 
break frequency outside the range of our data or a broken power law with no break, and the 
result is that both power law models had identical slopes. Arbitrarily forcing a specific break 
frequency produced fits to the data with low-frequency slopes of $\sim -0.3$ and high-frequency 
slopes of $\sim -1.4$, neither of which approximate the expected slopes of $-1$ and $-2$ seen in
most AGNs and Seyferts. From this we conclude that if $\nu_{B}$ is actually within the range 
detectable from our data, that our PSD is of such low quality that we are unable to measure it, 
or alternatively, $\nu_{B}$ lies at a frequency higher than $\sim 5 \times 10^{-3}$ Hz. 

\begin{figure}
\epsscale{1.0}
\plotone{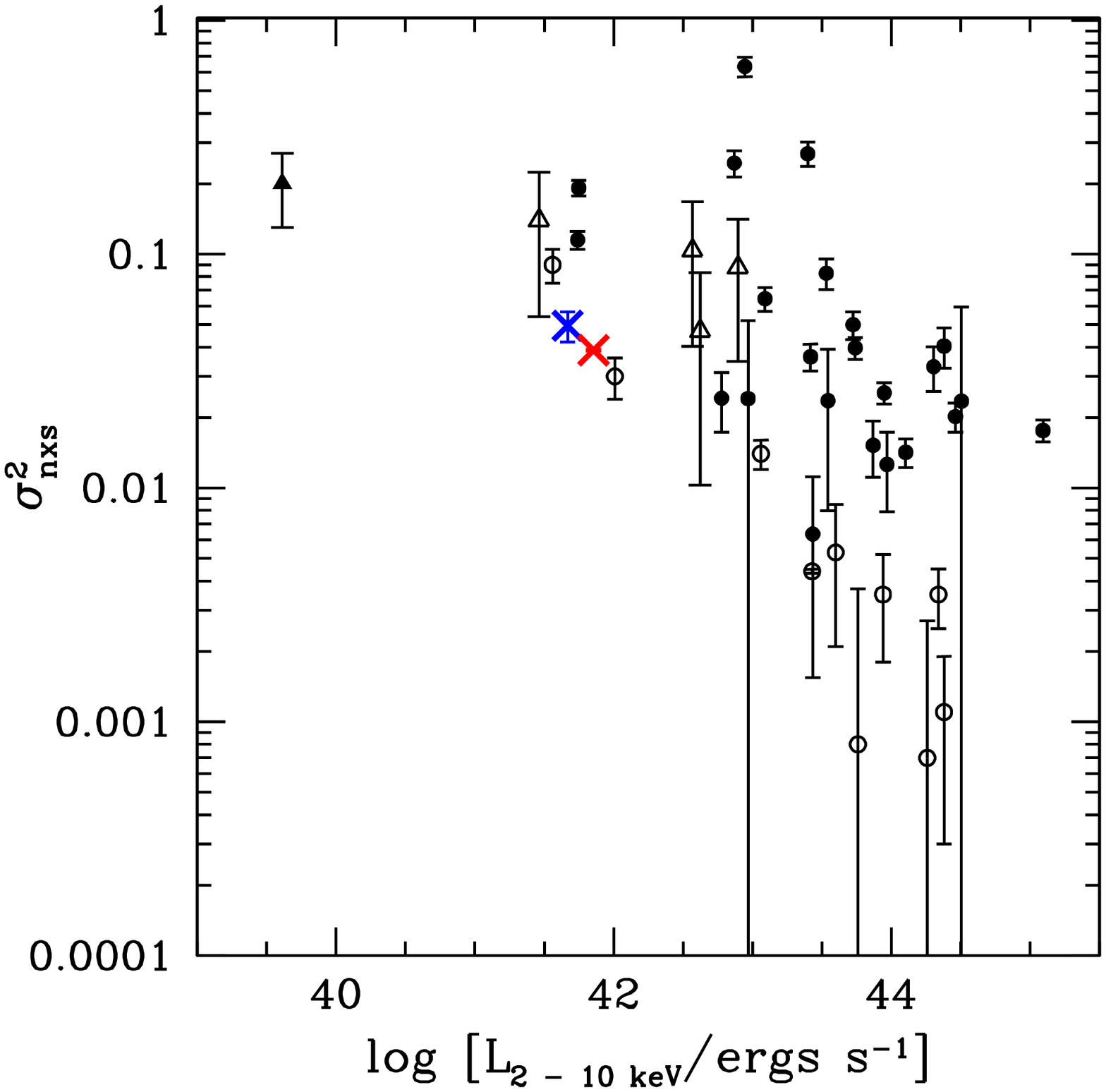}
\caption{Plot of $\sigma_{\rm nxs}^{2}$ vs. $2 - 10$ keV luminosity. NLS1 data from 
\citet{Leighly99a, Lei} are shown as filled circles. Classical Seyfert 1 data from 
\citet{Nandra, Nandra97b} are shown as open circles. POX 52 is shown as a red (\chandra\ data) 
or a blue (\xmmn\ data) cross. The unabsorbed luminosity is used for the \xmmn\ data point in 
order to represent the intrinsic luminosity of the source. Similar low-mass objects from the GH 
sample are shown as open triangles \citep{Mini}, including NGC 4395 \citep{Iwa00} as a filled 
triangle.}
\label{nxsplot}
\end{figure}

Even without an accurate measurement of $\nu_{B}$, we can still gain information from the 
relationship between excess variance and $L_{\rm X}$. There seems to be a noticeable split 
between the distribution of $\sigma^{2}_{\rm nxs}$ for NLS1 and classical Seyfert 1 galaxies, 
with the excess variance of the NLS1 galaxies consistently larger than that of the Seyfert 1 
galaxies at a given luminosity \citep{Leighly99a}. To test this for the low-mass case of POX 52,
we calculate the excess variance from the entirety of the usable light curve for both our \xmmn\
(second half of the observation, $\sim 50$ ks) and \chandra\ ($\sim 25$ ks) data, using 500 s 
and 100 s bins, respectively and plot the results against the $2 - 10$ keV unabsorbed 
luminosity, along with NLS1 data from \citet{Leighly99a, Lei}, Seyfert 1 data from 
\citet{Nandra, Nandra97b}, and GH object data from \citet{Mini} in Figure \ref{nxsplot}. From
this analysis, we find the excess variance of the two observations to be quite similar, 
$\sigma^{2}_{\rm nxs}(\chandra) = 0.0388 \pm 0.0003$ and 
$\sigma^{2}_{\rm nxs}(\xmmn) = 0.0494 \pm 0.0073$, despite the substantial additional absorber 
present in the \xmmn\ observation. This suggests that in the case of POX 52, the variability 
seen in the \xmmn\ observation is due to intrinsic variability and not due to fluctuations in 
the partial covering or the absorbing column density.

\begin{figure}
\begin{center}
\rotatebox{-90}{\scalebox{0.3}{\includegraphics{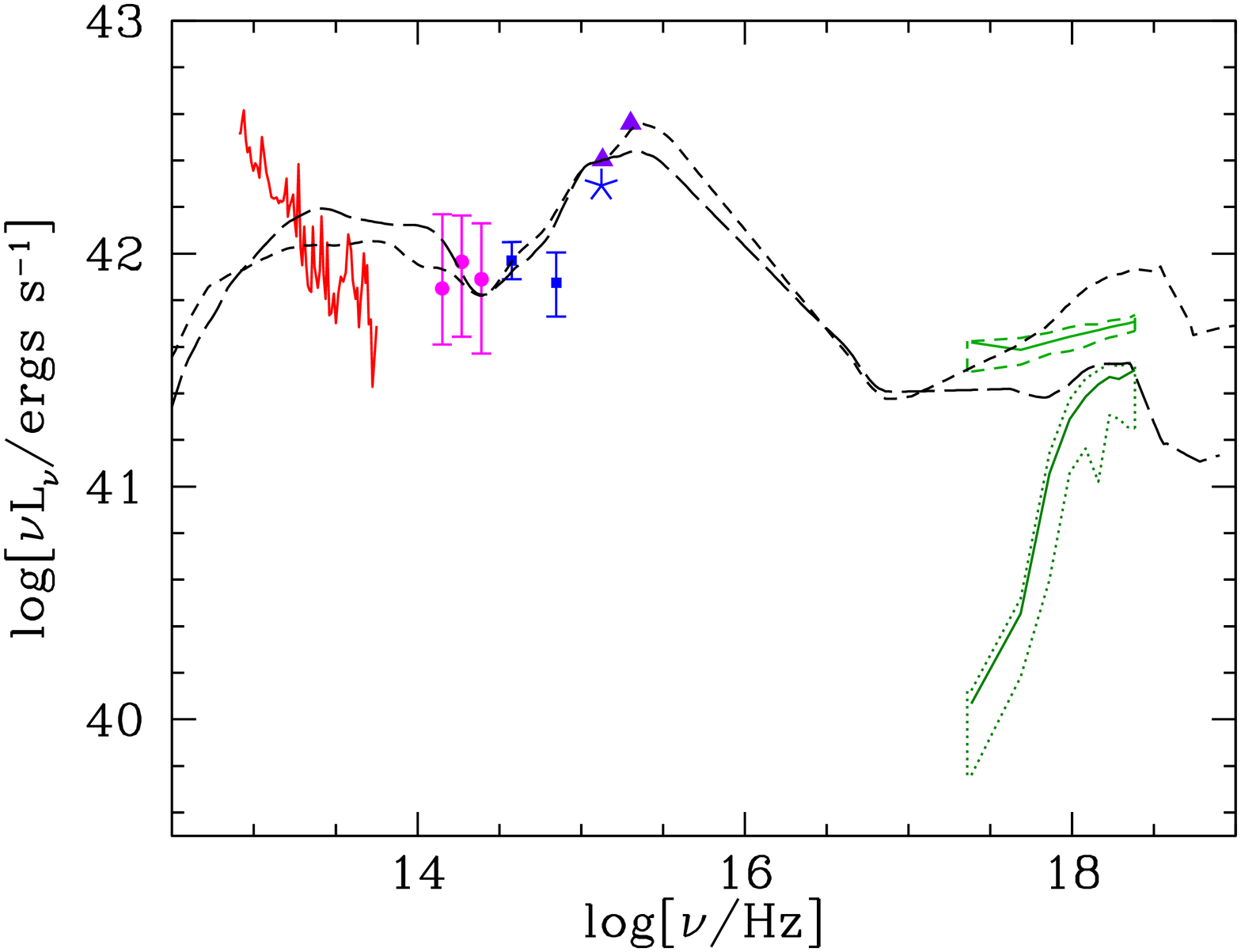}}}
\rotatebox{-90}{\scalebox{0.3}{\includegraphics{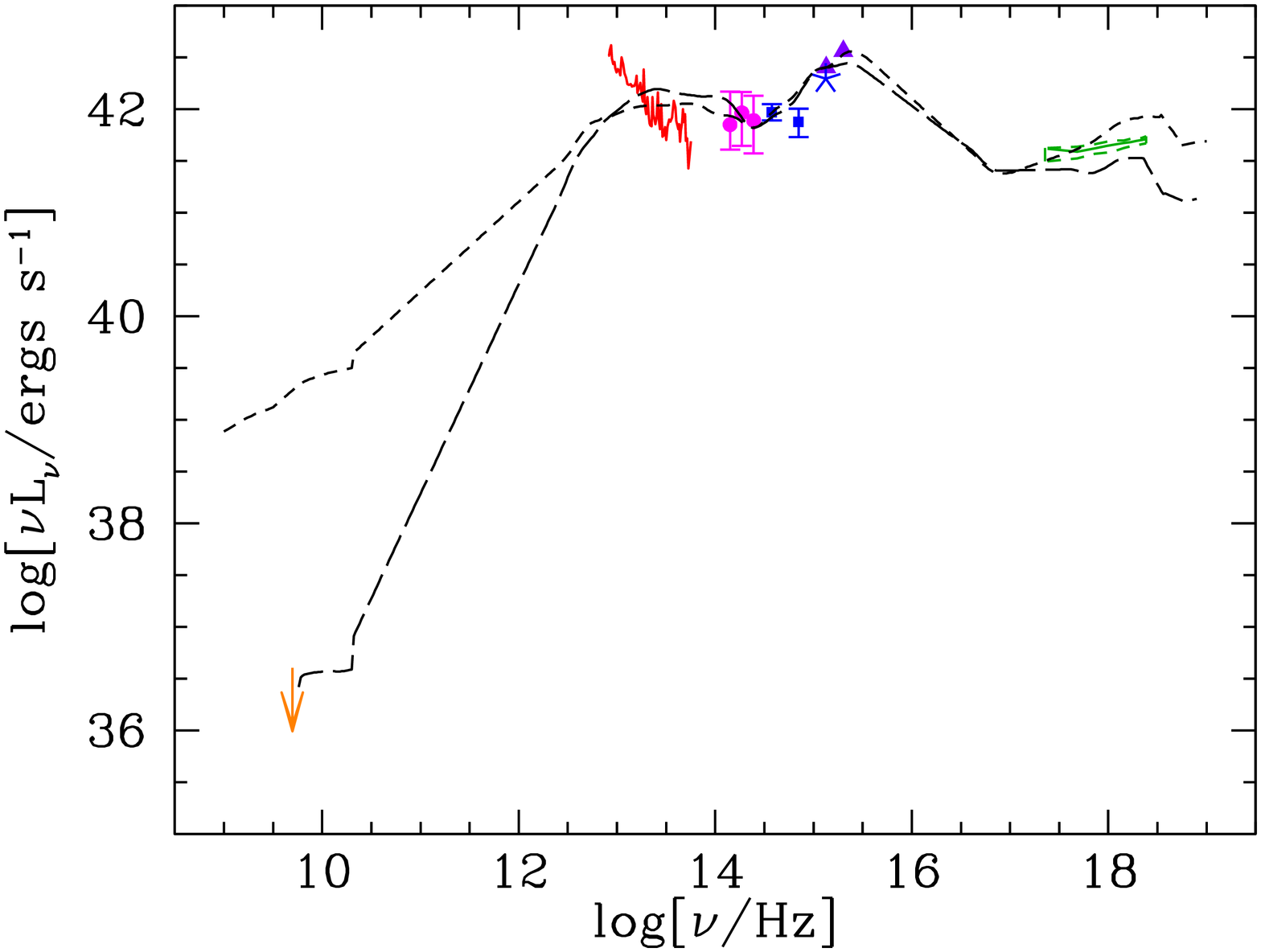}}}
\end{center}
\caption{SED of POX 52. The dashed lines are the radio-loud (short dash) and radio-quiet (long 
dash) SED templates from \citet{Elvis94} normalized to the UV data point. The X-ray data is in 
green with the dotted (\chandra) and dashed (\xmmn) lines representing the 1$\sigma$ uncertainty
range for the respective observatories. The \hst\ ({\it squares}) and OM ({\it asterisk}) data 
are in blue, the magenta circles are from 2MASS, the purple triangles are from \emph{GALEX}, the
red spectrum is from {\it Spitzer} IRS (binned by a factor of 5 for clarity), and the gold arrow
represents the upper limit on the VLA data.The {\it top panel} shows a close-up of the IR to 
X-ray data and the {\it bottom panel} shows the full extent of the data, excluding the strongly 
absorbed \xmmn\ observation.}
\label{SED}
\end{figure}

\subsection{Spectral Energy Distribution}
We combine all of our data and present one of the most complete SEDs (Figure \ref{SED}, Table 
\ref{SEDtable}) of a low-mass AGN to date, covering the radio to the X-ray. For comparison, we 
include the templates of \citet{Elvis94} from Palomar-Green (PG) quasars \citep{SG}, normalized 
to the flux of our NUV {\it GALEX} data point. The data from our observations closely match 
those of the radio-quiet (RQ) template for the majority of the frequency range shown. The 
\chandra\ data show POX 52 as X-ray bright compared to the RQ template and we again note that 
the \xmmn\ data represent a period of high absorption, not a decrease or change in the flux of 
the central object itself. It should also be noted that the observations are non-simultaneous 
and could therefore represent POX~52 in various states of emission, so care should be taken in 
examining the SED and comparing fluxes from different observations.

\begin{figure}
\epsscale{1.0}
\plotone{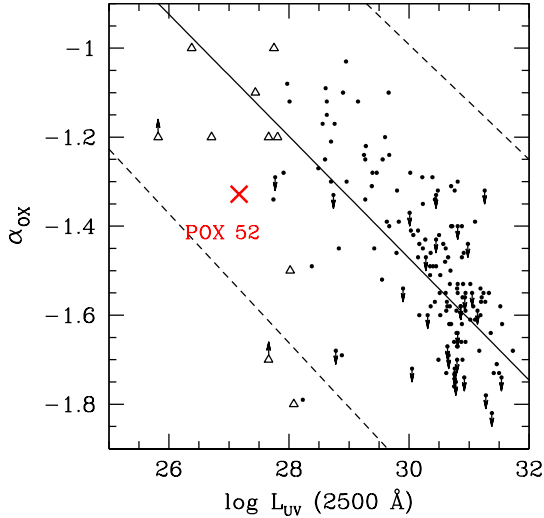}
\caption{Plot of \alOX\ vs.\ UV luminosity. Black dots are luminous type 1 AGNs from 
\citet{Str05}, arrows represent upper limits, the open triangles are intermediate-mass black 
holes from \citet{GH07a}, and the red cross is POX 52 derived from simultaneous UV and X-ray 
data taken with \xmmn. The solid line is the linear relation from \citet{Steffen} and the dashed
lines represent the $1\sigma$ error on that relation.}
\label{aOX}
\end{figure}

We approximate the shape of the SED near the OM observations as a power law with 
$f_{\nu} \propto\nu^{-1}$ and extract a UV luminosity at $2500~$\AA\ of 
$1.5 \times 10^{27}$~\ergsH. Combining this with the 2~keV unabsorbed luminosity derived from 
the simultaneous observations with the EPIC cameras on \xmmn, we can calculate \alOX, the slope 
of a hypothetical power law connecting the UV to the X-ray\footnotemark\ \citep{T79}. We find 
\alOX\ $= -1.3$, and this value does not change upon substitution of the \chandra\ 2 keV 
unabsorbed luminosity. \citet{Str05} studied \alOX\ in a sample of 228 optically selected AGNs 
and found that \alOX\ decreased with increasing luminosity. This correlation has been updated by
\citet{Steffen} to include higher redshift objects, resulting in 
$\alOX\ = -0.137L_{\rm UV} + 2.638$, where $L_{\rm UV}$ is the luminosity of the object 
measured at $2500$~\AA. We plot the data from Strateva \etal\ along with that of POX 52 and 
similar intermediate-mass black holes from \citet{GH07a} in Figure \ref{aOX}, showing that POX 
52 falls among its intermediate-mass counterparts and within the $1\sigma$ errors of the 
extrapolated correlation between \alOX\ and UV luminosity calibrated for more luminous type 1 
AGNs from Steffen \etal
\footnotetext{If a power law of the form $f_{\nu} \propto \nu^{\alOX}$ is assumed, \alOX\ is 
then defined by \alOX\ $= 0.3838 {\rm log}[f_{\nu}(2\ $keV$)]/[f_{\nu}(2500\ $\AA$)]$.}

\begin{deluxetable}{lll}
\tablewidth{3.0in}
\tablecaption{SED Data}
\tablehead{\colhead{Rest Wavelength} & \colhead{log [$\nu L_{\nu}$/ergs s$^{-1}$]} & 
\colhead{Source} \\
\colhead{or Energy} & \colhead{} & \colhead{}}
\startdata
6 cm       & $<$ 36.127  & VLA \\
36 \um     &  42.58  & \emph{Spitzer} IRS \\
30 \um     &  42.39  & \emph{Spitzer} IRS \\
20 \um     &  42.22  & \emph{Spitzer} IRS \\
10 \um     &  41.75  & \emph{Spitzer} IRS \\
6 \um      &  41.70  & \emph{Spitzer} IRS \\
2.12 \um   &  41.85  & 2MASS $K_S$ \\
1.61 \um   &  41.97  & 2MASS $H$ \\
1.22 \um   &  41.89  & 2MASS $J$ \\
7966 \AA   &  41.97  & \hst\ F814W \\
4257 \AA   &  41.88  & \hst\ F435W \\
2260 \AA   &  42.29  & \xmmn\ OM \\
2222 \AA   &  42.40  & \emph{GALEX} NUV \\
1495 \AA   &  42.56  & \emph{GALEX} FUV \\
1.0 keV    &  41.62  & \chandra \\
2.0 keV    &  41.59  & \chandra \\
10.0 keV   &  41.71  & \chandra \\
\enddata
\label{SEDtable}
\end{deluxetable}

Using the \citet{Elvis94} UV bolometric correction, we estimate \lbol\ 
$\approx~1.1~\times 10^{43}$~\ergs\ from the OM observations. Repeating this with the Elvis 
\etal\ $B$-band correction and the \hst\ measurement, results in 
\lbol\ $\approx 6.5 \times 10^{42}$\ \ergs, which is a factor of $\sim 2$ smaller than that of 
the UV data. Both of these estimates are based on observations of POX 52 in a single band, 
constituting a single point each on the SED. Using the templates derived by \citet{Elvis94}, 
shown on the SED, we estimate the \lbol\ using all of the data collected. With the templates 
normalized to match the {\it GALEX} NUV luminosity of POX 52, the radio-quiet template matches 
most data points at frequencies less than $10^{16}$ Hz, including the upper limit in the radio. 
We integrate the radio-quiet template from 6 cm to 1 keV and the \chandra\ data from 1 keV to 10
keV and estimate the bolometric luminosity based on the full extent of our data to be 
\lbol(SED) $\approx 1.3 \times 10^{43}$ \ergs, which is consistent with the value estimated from
the UV correction. We take this value as our best estimate of \lbol.

\subsection{Black Hole Mass}
For Seyfert 1 galaxies and quasars, there exists a correlation between the radius of the 
broad-line region (\rblr) and the luminosity of the optical non-stellar continuum (most often 
measured at 5100\AA), $\rblr \propto L^{\beta}$. \citet{Kaspi} derived this relationship based 
on their observations of 17 reverberation-mapped PG quasars combined with the results of 17 
lower-luminosity AGNs with reverberation data. This relationship was recently updated by 
\citet{bentz06} by taking into account the host galaxy starlight contamination. Using this 
relation in tandem with the virial relation, $\mbh~=~f~\rblr~\Delta~V^2/G$, one can estimate 
the mass of a black hole with a single spectrum. However, the normalization and slope in the 
radius-luminosity relation, and the value of the normalization factor $f$ in the virial 
relation, which depends on the geometry and kinematics of the broad line region, are still 
subject to considerable uncertainty. In addition, different permutations of this relation not 
only use different emission lines for the line width measurement $\Delta V$, but they also 
measure $\Delta V$ in different ways, either using the FWHM or the line dispersion 
($\sigma_{\rm line}$). \citet{onk04} calibrated the average normalization factor, $f = 1.4$ by 
assuming all AGNs follow the same \msigma\ relation as quiescent galaxies. 

We rely on the previous measurement of the \hbeta\ line width from a high-resolution Keck 
spectrum by \citet{bar04}. They fit the broad \hbeta\ line with a Gaussian model and found 
\vfwhm~$=~765$~\kms. The conversion from FWHM to $\sigma_{\rm line}$ for a Gaussian fit is 
$\sigma_{\rm line} =~$FWHM$/2.35$.

Using the radius-luminosity relation of \citet{Kaspi} and the broad \hbeta\ line width of 
\vfwhm~$=~765$~\kms\ discussed above, \citet{bar04} previously estimated the black hole mass of 
POX 52 to be \mbh~$\approx~1.6~\times~10^{5}$~\msun. We start our new calculation of black hole 
mass using the combined revisions of Bentz \etal\ and Onken \etal, and our new \hst\ 
measurements, such that
\begin{equation}
\mbh = 1.05 \times 10^7 \left(\frac{\lambda L_{\lambda} (5100~\mbox{\AA})}{10^{44}~{\rm \ergs}}
\right)^{0.518} \left(\frac{\vfwhm}{10^3~{\rm \kms}}\right)^2 \msun.
\end{equation}
We note that our \hst\ imaging is non-simultaneous with the previous Keck spectrum from Barth 
\etal, and therefore potentially introduces some error in the \mbh\ calculation due to source 
variability. The extinction-corrected point source absolute magnitude calculated from the GALFIT
decomposition is $M_{F435W}({\rm PSF}) = -15.7$. For comparison, we also calculate the 
extinction-corrected absolute magnitude needed to maximize the PSF contribution to the galaxy, 
$M_{F435W}({\rm PSF_{MAX}})~=~-16.3$. These two magnitudes correspond to a range of 
$\lambda L_{\lambda} (5100~$\AA$) = (3.2 - 5.6) \times 10^{41}$\ \ergs. 
Using \vfwhm\ $= 765$\ \kms\ from \citet{bar04}, we calculate \rblr~$= 2.0 - 2.7$ lt.\ days and 
\mbh~$\approx (3.1 - 4.2) \times 10^{5}$ \msun. 

As noted earlier, there are many different recipes to estimate black hole mass that rely on 
different versions of the radius-luminosity and virial mass relations. \citet{McGill} recently 
compiled a list of some of the most common versions and we use their formulations to explore the
range of possible \mbh\ values in POX~52. We eliminate those relations in which the slope of the
radius-luminosity relation is $\ge 5.5$, which contradicts the updated measurement of 
\citet{bentz06} and are left with three alternate relations (equations 8, 10 and 13 of McGill 
\etal). We use the same parameters above from our \hst\ measurements and the \vfwhm\ from 
\citet{bar04}, and find a range of black hole masses of 
\mbh~$\approx (1.8 - 2.7) \times 10^{5}$ \msun\  for POX 52, with an average of 
$\langle$\mbh$\rangle~= 2.2 \times 10^{5}$ \msun. This then extends our total range of possible 
black hole mass to $(1.8 - 4.2) \times 10^{5}$ \msun, increasing the evidence that POX 52 
contains an intermediate-mass black hole. This mass range is due to the scatter in using 
slightly different relations in the calculation of \mbh, but the systematic scatter in using any
single relation is $\sim 0.5$ dex \citep{onk04, V}.

\subsubsection{Black Hole Mass Estimates from X-ray Variability}
Both $\nu_{B}$ and excess variance have been shown to be roughly correlated with black hole mass
\citep{Papa}. We investigate both of these methods in order to independently confirm our mass
estimates from optical data. We first estimate the expected $\nu_{B}$ from the properties of the
AGN. \citet{Mark} fit a simple relation with $1/\nu_{B}$ measured in days, 
$1/\nu_{B} = \mbh /10^{6.5}\ \msun$, in which $\nu_{B}$ is solely dependent on black hole mass. 
This relation seems to hold quite well for normal Seyfert 1 galaxies, but it underestimates 
black hole masses calculated from $\nu_{B}$ for NLS1 galaxies, which have been known to have 
shorter break timescales (and therefore higher break frequencies) than classical Seyfert~1 
galaxies with similar black hole masses \citep{Papa}. In order to try to account for this 
disparity, \citet{McHardy} incorporated bolometric luminosity in the $\nu_{B}$-\mbh\ relation, 
in an effort to account for the higher accretion rates seen in NLS1 galaxies compared to the 
average Seyfert 1. Taking the inverse of $\nu_{B} = 1/T_{B}$, where $T_{B}$ is known as the 
break timescale, McHardy \etal\ found the best fit in the $T_{B}$-\mbh-\lbol\ plane for their 
sample of Seyfert 1, NLS1, and Galactic black holes to be 
${\rm log} T_B = 2.10 {\rm log} \mbh - 0.98 {\rm log} \lbol - 2.32$, where $T_B$ is measured in 
days, \mbh\ is measured in units of $10^6~\msun$, and \lbol\ is measured in units of 
$10^{44}$~\ergs. We use our average estimate of black hole mass and bolometric luminosity to 
calculate the corresponding $\nu_{B}$ from each of these relations and find 
$\nu_{B} \approx 1.2 \times 10^{-4}$ Hz using the Markowitz \etal\ prescription and 
$\nu_{B} \approx 3.8 \times 10^{-3}$ Hz using the McHardy \etal\ prescription. Since our PSD 
analysis was unable to determine $\nu_{B}$, we note that $\nu_{B}$ estimated from the Markowitz 
\etal\ relation occurs near the visual break seen in Figure \ref{PSD}, and $\nu_{B}$ estimated 
from the McHardy \etal\ relation occurs at the high-frequency end of this same plot.

Other recent work has explored alternative methods to correlate the variability in observations 
too short for PSD analysis with black hole mass. Niko{\l}ajuk \etal\ (2004, 2006) calibrated the
relationship between the excess variance of the variability in a given observation and the mass 
of the black hole. The excess variance must be sampled from the high-frequency end of the PSD, 
above $\nu_{B}$ where the shape of the PSD is described by $P(\nu) \propto \nu^{-2}$. The 
cornerstone of this method is the assumption that the amplitude of the PSD at $\nu_{B}$ 
multiplied by $\nu_{B}$ is constant for all objects and independent of black hole mass, 
{\it i.e.}\ $P(\nu_{B})\nu_{B} = {\rm constant}$. This has been shown to be 
roughly accurate for a handful of objects studied to date \citep{Mark, McHardy04, Papa}. Since 
the excess variance is simply the integral of the PSD, it can then be related to $\nu_{B}$ and 
consequently \mbh. This technique has only been tested on a few low-mass black holes in other 
AGNs, namely NGC 4395 and the NLS1 NGC 4051. 

Without a measurement of $\nu_{B}$ for POX~52, we cannot determine that we are sampling the 
appropriate section of the PSD, and therefore, we continue on with the caveat that the mass 
estimates calculated using this method are tenuous at best. The \chandra\ observation had a 
higher signal-to-noise ratio than the \xmmn\ and therefore had a better chance at producing a 
reasonable result. We are unable to probe timescales shorter than the break timescale predicted 
by the \citet{McHardy} relation, $T_{B} \approx 279$ s, so we continue with this method using 
the simpler \citet{Mark} relation which gives a wide range of break timescales 
($T_{B} \approx 5460 - 13,660$ s) corresponding to our full \mbh\ range. We tested the method 
using time bins of 100 s and 500 s and noted no significant difference between the two, except 
for the fact that using a smaller time bin allows us to divide the observation into a larger 
number of segments and therefore probe lower break timescales. We calculated black hole masses 
using segment lengths of $T \approx 2,775 - 12,500$ s with 100 s time bins and segments of 
length, $T_{B} \approx 6,250 - 12,500$ s with 500 s time bins, resulting in an average black 
hole mass of $\mbh = (3.9 \pm 1.7) \times 10^5$ \msun, where the errors represent the range of 
\mbh\ calculated for various $T_{B}$. If the break frequency of POX 52 is 
$\nu_B \approx 10^{-4}$ Hz or lower, then we are able to probe the region above this frequency 
and calculate a black hole mass of $\mbh \approx 3.9 \times 10^5$ \msun. But if the break 
frequency is higher, near the $\nu_B \approx 10^{-4}$ Hz predicted by the \citet{McHardy} 
relation, we are unable to explore the region of interest in frequency space and our black hole 
mass calculations are inconclusive.

\subsection{Accretion Luminosity}
Using the range of black hole masses we have estimated here, the Eddington luminosity of POX 52 
is in the range \ledd\ $= (2.9 - 5.5) \times 10 ^{43}$ \ergs.  Taking the bolometric luminosity 
calculated from the SED, we find an estimate of the Eddington ratio of 
\lbol/\ledd\ $= 0.2 - 0.5$. 

NLS1 galaxies have been shown to have higher Eddington ratios on average than their Seyfert 1 
counterparts, usually with $L/\ledd \approx 1$ for NLS1 galaxies compared to 
$L/\ledd \approx 0.1$ for a normal Seyfert 1 \citep{Xu}. However, there are a few objects, e.g. 
NGC 4395, that strongly contradict this dichotomy. Despite nearly identical spectra in the 
optical, POX~52 has an Eddington ratio that is more than two orders of magnitude larger than the
\lbol/\ledd\ $= 10^{-3}$ for NGC 4395 \citep{Pet05}, demonstrating that there must be 
substantial differences between the central engines of these prototypical low-mass AGNs. An 
Eddington ratio of $\sim 0.5$ suggests POX~52 is not as highly accreting as many NLS1 galaxies 
and may represent an intermediate object in the continuum between NLS1 and Seyfert 1 galaxies. 

The unabsorbed spectral slope ($\Gamma$) of POX 52 remained relatively unchanged between X-ray 
observations, suggesting little change in the intrinsic continuum emission. \citet{Shemmer06} 
studied the correlation between X-ray spectral slope and accretion rate (\lbol/\ledd) for AGNs
up to $z \approx 2$, increasing the luminosity range to more than 3 orders of magnitude, almost 
twice as large as previous studies (e.g. Porquet \etal\ 2004, Wang \etal\ 2004, Bian 2005). 
Despite having a much lower mass than any of sources used in the Shemmer \etal\ study 
(\mbh $\approx 10^7 - 10^{10}$ \msun), POX 52 falls among the majority of objects in the sample 
when plotting $\Gamma$ versus \lbol/\ledd, making it consistent with this correlation, and
suggests this relationship is independent of black hole mass. With $\Gamma =1.8$ and 
\lbol/\ledd\ $\approx 0.2 - 0.5$, POX 52 falls in the overlapping region of the apparent 
distribution split between NLS1 ($\Gamma \approx 1.75 - 2.6$, \lbol/\ledd\ $\approx 0.5 - 1.5$) 
and Seyfert 1 ($\Gamma \approx 1.5 - 2.3$, \lbol/\ledd\ $\approx 0.05 - 0.5$) galaxies. It is 
important to note that bolometric luminosity is most often estimated from a single-band using a 
bolometric correction (as done in Shemmer \etal), and that there might be significantly more 
scatter or bias in these relations because of these corrections. However, since \lbol\ for POX 
52 is calculated using a compilation of datasets over a wide range of wavelengths, it lends 
itself to a more secure calculation of the accretion luminosity of this object.

\subsection{Black Hole -- Host Galaxy Correlations}
Relationships connecting the black hole mass with host galaxy properties, such as velocity 
dispersion, bulge mass, and bulge luminosity can be useful in predicting the mass of a black 
hole of many objects from single photometric or spectroscopic observations. Determining the 
slope and amount of scatter in these correlations, specifically in the low-mass and high-mass 
ends of the black hole mass spectrum, can help us learn how black holes and host galaxies, 
evolve together in time (e.g. di Matteo \etal\ 2005).

If we extrapolate the \msigma\ relation of \citet{Tre02} down to $\sigmastar = 36$~\kms\ 
\citep{bar04}, we find a black hole mass of \mbh~$=~(1.4~\pm~1.1)~\times~10^{5}$~\msun, which is
consistent with the average black hole mass calculated from the mass estimators described above.
We use the F435W magnitude of the outermost \sersic\ component in order to estimate an absolute 
$B$-band magnitude for the host galaxy of $M_{B} \approx -16.0$. Extrapolating the \citet{MH03} 
\mlbul\ relation down to a lower luminosity, we find a predicted \mbh\ of 
\mbh $\approx 8.7 \times 10^{5}$ \msun, which is a factor of $2-5$ higher than current 
estimates. The updated version of the \citet{MH03} relation from \citet{Graham} predicts a 
black hole mass of \mbh $\approx 7.4 \times 10^{6}$ \msun, which is a factor of $20-40$ larger 
than estimated here.  

The measured host galaxy color places a constraint on the age (and thereby mass-to-light ratio) 
of the galaxy stellar populations. The code {\it kcorrect} \citep{Blanton} fits a 
combination of templates with a range of star formation histories, metallicities, and
reddening values and returns the stellar mass [assuming a Chabrier (2003) initial mass 
function].  Using either the \hst\ measurements presented here, or the $BVRI$ measurements from 
Barth \etal\ (2004), we find a stellar mass of $M_{\rm host} \approx 1.2 \times 10^{9}$ \msun. 
If we extrapolate the \citet{HR} \ensuremath{\mbh-\mbul} relation to the host galaxy mass of 
POX~52, the bulge mass implies a black hole mass of \mbh~$\approx 1.1 \times 10^{6}$~\msun, a 
factor of $2 - 6$ larger than our current estimate. 

The compact systems investigated in \citet{GH08} follow the same trends. For the \mlbul\ 
relation, the black hole masses are on average two orders of magnitude less massive than their 
host galaxy luminosities would predict and are similarly an order of magnitude less massive than
their host galaxy masses would predict. The divergent \ensuremath{\mbh-\mbul} and \mlbul\ 
relations in these galaxies as compared to elliptical galaxies may be an outcome of their 
different formation histories, as reflected in their different fundamental plane locations, 
which still leaves the possibility that the \msigma\ relation represents a tighter empirical
correlation in this low-mass regime. The \msigma\ relation seems to more accurately reflect the 
properties of lower mass black holes and their host galaxies \citep{bar05}, whereas the \mlbul\ 
and \ensuremath{\mbh-\mbul} relations both overestimate the black hole mass by factors of 
$\sim 2-100$. 

There has been recent work suggesting a correlation between black hole mass and \sersic\ index 
for elliptical galaxies or the bulges of disk galaxies, for the purpose of predicting black hole
masses in other objects \citep{GD07}. The measured \sersic\ index $n \approx 4$ and mass \mbh\ 
$\approx 3 \times 10^{5}$ \msun\ make POX 52 a strong outlier in this relation, which previously
only included objects with \mbh\ $\approx 10^{6} - 10^{9}$ \msun. For $n = 4$, the mass 
predicted by this relation is \mbh\ $\approx 10^{8}$ \msun, approximately 3 orders of magnitude 
larger than all other mass estimates of POX 52 suggest. The inclusion of more objects like 
POX~52 and those in the GH sample demonstrate that there cannot be a single, tight \mbh$-n$
correlation applicable across the entire mass spectrum of galaxies. 

Recent \hst\ imaging surveys have found evidence for a transition in the population of central 
star-cluster nuclei in early-type galaxies, such that galaxies more luminous than 
$M_B \approx -20.5$ mag, or more massive than a few $\times 10^{10}$ \msun, tend to lack 
resolved central nuclei, while lower-luminosity early-type galaxies are more likely to be 
nucleated \citep{Wehner06, Ferr06, Cote06}. These groups further found that the black holes in 
luminous ellipticals and the stellar nuclei in lower-mass galaxies followed the same 
relationship between central object mass and host galaxy mass. They suggested that formation of 
a central massive object might be a universal property of early-type galaxies, but that the 
central object would be \emph{either} a massive black hole or a central star cluster, depending 
on the host galaxy mass. It has already been demonstrated that NGC 4395 has both an AGN and a 
nuclear star cluster \citep{Matthews} and our GALFIT decompositions suggest the possibility that
POX 52 might as well. Observations such as these provide further motivation to test for the 
presence of central black holes in dwarf ellipticals through dynamical searches or AGN surveys, 
and POX 52 is a unique object in this context, given its low stellar luminosity 
($M_B = -16.0$ mag) and the unambiguous presence of a central black hole.

\section{Summary and Conclusions}
New \hst\ observations have confirmed POX 52 to be a dwarf elliptical galaxy, with a \sersic\ 
index of $n \approx 4.3$, $r_e \approx 1.15$, and $M_I = -17.3$. These parameters, along with 
the previously measured stellar velocity dispersion from \citet{bar04}, show POX 52 to reside 
near the dwarf elliptical family in the fundamental plane. POX 52 was the first AGN found 
to reside in a dwarf elliptical galaxy, but more recent studies of low-mass AGNs and their host
galaxies have found a number of similar objects \citep{GH08}. 

We see strong variability over $500 - 10^{4}$ s timescales with both \chandra\ and \xmmn\ 
and a substantial change in the absorbing column density due to partial covering over a 9 month 
time period. Data from the Optical Monitor on \xmmn\ showed no significant variability in the UV
over short $\sim 10^3 - 10^5$ s timescales. We see no variability within the $2\sigma$ 
errors in the 7 month period between the OM and {\it GALEX} NUV observations, but this does 
allow variations of $10-35$\% on the $1\sigma$ uncertainty level. The {\it Spitzer} spectrum
shows evidence of possible star-formation due to the PAH features and red continuum, but that 
analysis is deferred to a later paper.

Including data from the VLA, {\it Spitzer}, \emph{GALEX}, and 2MASS, we compile an extensive 
SED that shows POX 52 to be consistent with the radio-quiet template of more massive quasars and
estimate a bolometric luminosity of \lbol\ $= 1.3 \times 10^{43}$~\ergs\ by incorporating all 
of the available data. Using a variety of mass estimators, we find a range of black hole mass 
\mbh~$\approx (1.8-4.2) \times 10^{5}$ \msun. We find that we are unable to determine the break 
frequency from the \chandra\ PSD, either due to a low signal-to-noise ratio or insufficient 
temporal sampling. If $\nu_B \approx 10^{-4}$ Hz, then our mass estimate from the X-ray excess 
variance of $\mbh = (3.9 \pm 1.7) \times 10^5$ \msun\ is consistent with the optical mass 
estimates. Otherwise, new X-ray observations of POX 52 in an unabsorbed state with the larger 
effective area of \xmmn\ would be needed in order to determine $\nu_B$ and estimate the black 
hole mass with this method.

We have shown POX 52 to have an SED consistent with a scaled-down version of more massive type 1
quasars and that it is in a different accretion state than NGC 4395, as seen in the two orders 
of magnitude disparity between Eddington ratios. The number of low-mass AGNs known is increasing
with new results from large AGN surveys, such as \citet{GH04, GH07c}, but POX 52 remains unique 
in that it is one of the nearest and best studied of the low-mass AGNs.

\acknowledgments 
We thank J. S. Ulvestad for his contribution to the VLA observations and C. Y.
Peng for his assistance with GALFIT. C.E.T. 
would like to thank M. Bentz, P. Humphrey, and M. Hood for numerous comments, discussions, and 
suggestions. Support for \hst\ program \#10239 was provided by NASA through a grant from the 
Space Telescope Science Institute, which is operated by the Association of Universities for 
Research in Astronomy, Inc., under NASA contract NAS 5-26555. Some of the data presented in this
paper were obtained from the Multimission Archive at the Space Telescope Science Institute 
(MAST). Support for MAST for non-\hst\ data is provided by the NASA Office of Space Science via 
grant NAG5-7584 and by other grants and contracts. Support for the analysis of \chandra\ data 
was provided by the National Aeronautics and Space Administration through \chandra\ Award Number
G05-6119X issued by the Chandra X-ray Observatory Center, which is operated by the Smithsonian 
Astrophysical Observatory for and on behalf of the National Aeronautics and Space Administration
under contract NAS 8-03060. The analysis of \xmmn\ data presented herein was supported by grant 
NNG05GQ33G from NASA. This work is based in part on observations made with the Spitzer Space 
Telescope, which is operated by the Jet Propulsion Laboratory, California Institute of 
Technology under a contract with NASA. Support for this work was provided by NASA through an 
award issued by JPL/Caltech. The IRS was a collaborative venture between Cornell University and 
Ball Aerospace Corporation funded by NASA through the Jet Propulsion Laboratory and Ames 
Research Center. This work was also supported by the National Science Foundation under Grant No.
AST-0548198. This publication makes use of data products from the Two Micron All Sky Survey, 
which is a joint project of the University of Massachusetts and the Infrared Processing and 
Analysis Center/California Institute of Technology, funded by the National Aeronautics and Space
Administration and the National Science Foundation.

\end{document}